\documentclass[namedreferences]{solarphysics}
\usepackage[optionalrh,natbib]{spr-sola-addons} 
\usepackage{graphicx}        
\usepackage{courier}         
\usepackage{bm}
\usepackage[usenames,dvipsnames]{color}             
\usepackage{url}             
\usepackage{tabularx,ragged2e,booktabs,caption} 
\usepackage{longtable} 
\newcommand{\sign}{\mathop{\mathrm{sign}}}

\newcommand{\rmd}{ {\mathrm d} }
\renewcommand{\vec}[1]{\bm{#1}}

\newcommand{\degree}{^\circ} 

\newcommand{\BE}{\begin{equation}}
\newcommand{\EE}{\end{equation}}
\newcommand{\BA}{\begin{eqnarray}}
\newcommand{\EA}{\end{eqnarray}}
 \newcommand{\fig}[1]{Figure~\ref{fig:#1}}
 \newcommand{\figs}[2]{Figures~\ref{fig:#1} and \ref{fig:#2}}
 \newcommand{\figsss}[4]{Figures~\ref{fig:#1}{#2} and \ref{fig:#3}{#4}}
 \newcommand{\figss}[2]{Figures~\ref{fig:#1}\,--\,\ref{fig:#2}}
 \newcommand{\sect}[1]{Section~\ref{sec:#1}}

\newcolumntype{C}[1]{>{\Centering}m{#1}}

\newcommand{\eg}{\textit{e.g.}}
\newcommand{\etal}{\textit{et al.}}
\newcommand{\ie}{\textit{i.e.}}

\newcommand{\Nt}{N_{\rm t}}

\newcommand{\Ntc}{N_{\rm t,c}}
\newcommand{\NtcMean}{\overline{N_{\rm t,c}}}
\newcommand{\NtcMax}{N_{\rm t,c,max}}
\newcommand{\phiMean}{\overline{\phi}}

\newcommand{\tauMean}{\overline{\tau}}
\newcommand{\tauMax}{\tau_{\rm max}}
\newcommand{\tauc}{\tau_{\rm c}}
\newcommand{\taucMean}{\overline{\tau_{\rm c}}}
\newcommand{\taucMax}{\tau_{\rm c,max}}
\newcommand{\xs}{x_{\rm s}}
\newcommand{\ys}{y_{\rm s}}

\begin{document}

\begin{article}

\begin{opening}

\title{Evidence of Twisted flux-tube Emergence in Active Regions}

\author{M.~\surname{Poisson}$^{1}$ \sep 
C.H.~\surname{Mandrini}$^{1,2}$\sep
P.~\surname{D\'emoulin}$^{3}$ \sep 
M.~\surname{L\'opez Fuentes}$^{1,2}$}

\runningauthor{M. Poisson \etal }
\runningtitle{Magnetic Tongues in Emerging Active Regions}

\institute{$^{1}$ Instituto de Astronom\'\i a y F\'\i sica del Espacio (IAFE), CONICET-UBA, Buenos Aires, Argentina\\
Corresponding author M. Poisson, \\ email: \url{marianopoisson@gmail.com}\\             
             $^{2}$ Facultad de Ciencias Exactas y Naturales (FCEN), UBA, Buenos Aires, Argentina\\
             $^{3}$ Observatoire de Paris, LESIA, UMR 8109 (CNRS), F-92195 Meudon Principal Cedex, France}

\begin{abstract}
Elongated magnetic polarities are observed during the emergence phase of bipolar active regions (ARs). 
These extended features, called magnetic ``tongues'', are interpreted as a consequence of the azimuthal component of the magnetic flux in the toroidal flux-tubes that form ARs.
We develop a new systematic and user-independent method to identify AR tongues. Our method is based on the determination and analysis of the evolution of the AR main polarity inversion line (PIL). The effect of the tongues is quantified by measuring the acute angle [$\tau$] between  the orientation of the PIL and the direction orthogonal to the AR main bipolar axis. We apply a simple model to simulate the emergence of a bipolar AR. This model lets us interpret the effect of magnetic tongues on parameters that characterize  ARs ({\it e.g.} the PIL inclination, and the tilt angles and their evolution). In this, idealized kinematic emergence model, $\tau$ is a monotonically increasing function of the twist and has the same sign as the magnetic helicity.
We systematically apply our procedure to a set of bipolar ARs (41 ARs) that were observed emerging in line-of-sight magnetograms over eight years.
For the majority of the cases studied, the presence of tongues has a small influence on the AR tilt angle since tongues have much lower magnetic flux than the more concentrated main polarities. 
 From the observed evolution of $\tau$, corrected by the temporal evolution of the tilt angle and its final value when the AR is fully emerged, we estimate the average number of turns present in the sub-photospheric emerging flux-rope. These values for the 41 observed ARs, except one, are below unity. This indicates that sub-photospheric flux-ropes typically have a low amount of twist, {\it i.e.} highly twisted flux-tubes are rare. Our results demonstrate that the evolution of the PIL is a robust indicator of the presence of tongues and constrains the amount of twist present in emerging flux-tubes.
\end{abstract}


\end{opening}


\section{Introduction}
 \label{sec:introduction} 

It is widely recognized that magnetic helicity is one of the most relevant properties of the magnetic structures in the Sun. The presence of a significant amount of magnetic helicity has been confirmed in countless observed solar features such as sunspot whorls, sigmoids, prominences, coronal mass ejections (CMEs), and, also, in interplanetary magnetic clouds (see examples in \opencite{Demoulin09}). The magnetic-helicity content of active regions (ARs) is associated with the free magnetic energy available to be released by reconnection during flares and CMEs (see, \eg,  \opencite{Kusano04}). Being a conserved quantity in ideal magneto-hydrodynamic (MHD) regimes \cite{Berger84}, magnetic helicity provides information about the mechanisms of formation and evolution of magnetic fields in the solar interior and atmosphere, which can be traced from the convective zone to the corona and the interplanetary medium (see \opencite{Nakwacki11}, and references therein). 

Both numerical simulations and observations indicate that magnetic-flux-tubes that emerge from the solar interior and form active regions (ARs) transport magnetic helicity acquired when they develop, most likely at the bottom of the convective zone (\opencite{Fan09r}). 
Their magnetic helicity is due to the deformation of the magnetic-field lines around the main axis of the tube (the so-called twist) and the deformation of the axis as a whole (known as writhe: see \opencite{Lopez-Fuentes03}). flux-tubes with a non-null magnetic helicity are called flux-ropes. Ever since the first two-dimensional (2D) simulations (see, \eg , \opencite{Emonet98}), it has been proposed that, in order to survive the interaction with the convective zone plasma, the emerging flux-ropes must have a certain minimum amount of twist. Although more sophisticated 3D simulations provide different thresholds, twist is regarded as a fundamental feature of flux-tube emergence (\opencite{Fan01}).

It has been shown recently that twist has an important role during the crossing of the emerging structure through the sub-adiabatic interface between the top layer of the convective zone and the photosphere (see the review by \opencite{Hood12}). In the simulations, the top of the flux-rope expands and becomes flat as it ``crashes'' against the overlying photosphere since the ambient plasma pressure decreases very rapidly with height (\ie\ the scale height is about 150 km). Then, the flux-rope top is no longer buoyant and the magnetic flux is stored below the photosphere. When enough magnetic flux has accumulated, 
the so-called Parker instability develops and allows the magnetic flux to emerge into the solar chromosphere and corona. 
At the beginning, only small portions of the original flux-rope emerge in the form of thin sea-serpent tubes (\opencite{Strous96}; \opencite{Pariat04}). Recently, \inlinecite{Valori12} studied the emergence of an AR and found that the first manifestations of the emergence occur in the form of small bipoles, which are consistent with the sea-serpent scenario. The characteristic size of the bipoles can be associated with the instability wavelength (see also \opencite{Bernasconi02}; \opencite{Otsuji11}; \opencite{Vargas12}). As the flux continues to emerge in this way, coalescence and reconnection leads to the formation of larger flux concentrations and the observation of pores, sunspots, and faculae (see \opencite{Toriumi12}, and references therein). \citeauthor{Murray06} (\citeyear{Murray06}, \citeyear{Murray08}) showed that the amount of twist in the flux-rope can determine its ability to emerge into the atmosphere. The higher the twist and the stronger the magnetic tension are, the greater the tube cohesion is and the faster the instability and emergence occur. Numerical studies predict that the emergence process leads to the formation of a new flux-rope and the development of a sigmoidal structure in the corona (see, \eg,\  \opencite{Hood09}).

The structure of flux-ropes determines many of the features observed in line-of-sight photospheric magnetograms during flux emergence. Among them, one of the most conspicuous is the presence of magnetic ``tongues'' (or ``tails''). These features are observed during the emergence of the top part of ``$\Omega$-shaped'' flux-ropes and are produced by the line-of-sight projection of the azimuthal component of the magnetic field.
Magnetic tongues are observed in magnetograms as deformations or extensions of the main polarities that form bipolar ARs. They were originally reported by \inlinecite{Lopez-Fuentes00} and observed later in other works 
(see, \eg,\ \opencite{Li07}; \opencite{Chandra09}; \opencite{Luoni11}). Their presence in emerging flux-ropes has been found by numerical simulations (\opencite{Archontis10}; \opencite{MacTaggart11}; \opencite{Jouve13}). \inlinecite{Luoni11} compared the helicity-sign inferred from the orientation of magnetic tongues, observed in line-of-sight magnetograms, with other known helicity sign proxies for a sample of 40 ARs. To characterize the effect of the tongues, these authors computed the elongation of the main AR polarities and the evolution of the polarity inversion line (PIL). 
Their results confirmed that the photospheric magnetic-flux distribution, associated with the presence of tongues, can be used as a reliable proxy for the helicity sign. 

In this article we develop a novel systematic method to quantify the effect of the magnetic tongues during the emergence of ARs. Preliminary results of the application of our method have been given by \inlinecite{Poisson12}. In \sect{data}, we briefly describe the observations used in our study. We define a new parameter [$\tau$] (\sect{method}) to characterize the presence of tongues. $\tau$ measures the angular difference between the PIL orientation and the direction orthogonal to the main AR bipolar axis. We discuss our procedure to compute the inclination of an AR PIL (\sect{pil}) and its uncertainty, which is reflected in the confidence interval when we determine $\tau$ (\sect{tau-error}).  In \sect{results}, we systematically analyze all of the isolated bipolar ARs, observed during eight years, for which the full emergence was visible in line-of-sight magnetograms. We illustrate our results showing in detail a few examples of the application of our algorithm. These are compared with those coming from synthetic ARs obtained simulating the kinematic emergence of toroidal flux-ropes. In \sect{Global} we define and use the average and maximum values of the computed parameters to compare observed ARs having different characteristics. 
Finally, in \sect{conclusions}, we summarize our findings and conclude that the proposed systematic method provides a solid way to determine quantitatively the properties of magnetic tongues setting constraints on emergence models. 


\section{Observations and Data Analysis} 
\label{sec:data}  

The data used in this work come mainly from the {\it Michelson Doppler Imager} (MDI: \opencite{Scherrer95}), onboard the {\it Solar and Heliospheric Observatory} (SOHO). MDI provides full-disk maps of the line-of-sight component of the magnetic field in the photosphere with a spatial resolution of $\approx$ 1.98'. We use level 1.8 maps, which are the average of five magnetograms with a cadence of 30 seconds. These maps are constructed once every 96 minutes. The error in the flux densities per pixel in these MDI magnetograms is 9 G. For ARs appearing on the solar disk by the end of 2010 and later, when MDI provided magnetic data sporadically, we use observations by the {\it Helioseismic and Magnetic Imager} (HMI: \opencite{Schou12}), onboard the {\it Solar Dynamics Observatory} (SDO). HMI full-disk magnetograms have a spatial resolution of 0.5' and a 45 second cadence. The precision in the flux density per pixel in HMI line-of-sight measurements is 10 G. 

We select ARs that appeared on the solar disk between the start of the decay of Solar Cycle 23 ($\approx$ January 2003) and the beginning of the maximum of Cycle 24 (around the beginning of 2011). This solar minimum was peculiar in that it showed the lowest number of sunspots recorded in the last 100 years \citep{Bisoi14}. This is particularly favorable for this study, since isolated ARs with simple magnetic configurations and low background flux are best suited to derive AR properties. We systematically searched for ARs in this period of time (eight years) by examining SOHO/MDI and SDO/HMI images provided by the Helioviewer website (\url{www.helioviewer.org}). After selecting the ARs in this way, we retrieved the corresponding original data from the instrument data bases. 

We restrict our selection to bipolar ARs because in these cases the main PIL can be better defined and the magnetic tongues are not distorted by multiple flux emergences.  We choose ARs that started emerging before crossing the central meridian and not farther than 30$^{\circ}$ East from this location. In this way, most of their emergence phase can be observed close to disk center and foreshortening effects are minimized.
During this solar-minimum period, we find 41 well-isolated ARs emerging in the quiet Sun that fulfill the required bipolar configuration and emergence conditions. Although the constraints on the characteristics of the selected ARs limit the statistical strength of our results, the main aim here is to apply and test new methods to determine the degree of twist of ARs.

To build our set of analysis, we first rotate all full-disk magnetograms to the date of central meridian passage (CMP) of the region of interest.  In this way, we account for the known rotation rate of the Sun at the corresponding latitude.  We also apply corrections for the flux-density reduction due to the angle between the observer and the assumed radial magnetic-field direction (see \opencite{Green03} for a description of this radializing procedure). After this correction, the term PIL corresponds to the polarity inversion line of the component of the field normal to the photosphere. When radializing the observed magnetic field, we are neglecting the contribution of the field components in the photospheric plane; this contribution can be important when the AR lies farther from disk center but this is not the case for our set of selected ARs. For these first steps, we use standard Solar Software tools.

Then, from the set of full-disk magnetograms covering the evolution of each AR, we choose and extract rectangular boxes of variable sizes encompassing the AR main polarities. The selection is done by eye on the first, middle, and last magnetogram included in the analyzed period. The rectangles for the intermediate magnetograms are chosen by interpolating linearly between the locations of the corners of the first and middle rectangle, and from this one to the last chosen rectangle. We produce movies for each AR to check by eye that the selection done in this way contains all of the AR emerging flux within our rectangles. If this is not the case, we refine our procedure taking intermediate magnetic maps for the interpolation. The analyzed temporal range extends from the first appearance of the AR polarities until the maximum flux is reached and/or foreshortening effects are important.


\section{Characterizing Magnetic Tongues}
\label{sec:method}

\subsection{Why Use the Inclination of the PIL?}
\label{sec:aim}

 To illustrate the main characteristics of magnetic tongues,
in particular their link to the flux-rope twist, we first present a simplified model of the emergence of an AR. This model also shows how the PIL inclination lets us characterize the evolution of the magnetic tongues during the emergence phase. 
 
\begin{figure}[t]
\begin{center}
\includegraphics[width=.99\textwidth]{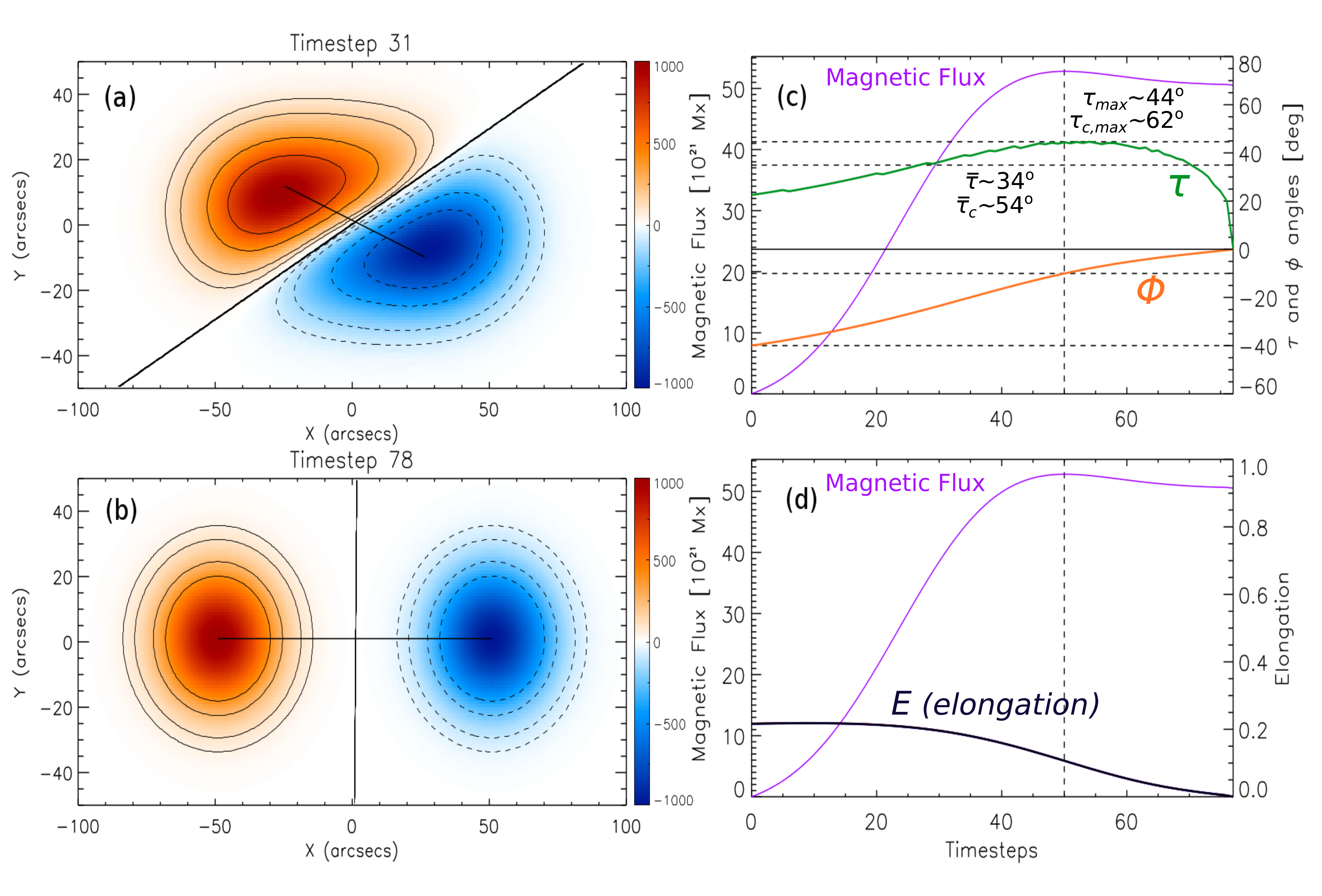} 
\caption{Synthetic magnetograms obtained from the AR-emergence model based on the rising of a toroidal flux rope having a positive twist, $\Nt = 1$, and an aspect ratio $a/R = 0.4$. (a) and (b) correspond to two different time steps of the simulated emergence.  The flux-rope axis is in the $x$-direction.
The continuous (dashed) lines are isocontours of the positive (negative) magnetic field component normal to the photosphere, drawn for four different values taken at 10, 25, 50, and 70\% of the axial field strength $B_{\rm max} = 2000$ G. The red and blue shaded-regions indicate positive and negative polarities, respectively.  
(c) and (d) illustrate the evolution of the magnetic flux (violet continuous line), $\tau$ (green line), $\phi$ (orange line) and the elongation for both polarities (black line) for the modeled AR.
The angles $\tau$ and $\phi$ are defined in Figure~2. Panel (c) shows the values of $\tauMean$,  $\tauMax$, $\taucMean$, and $\taucMax$; these parameters are defined in \sect{Global}. The vertical dashed line is drawn at maximum flux. The horizontal dashed lines, close to the corresponding curves, indicate particular values of $\tau$ and $\phi$.}
\label{fig:modN1}
\end{center}
\end{figure}

Let us mimic an AR emergence across the photosphere using a toroidal magnetic flux-tube with twist. The model is the same as the one presented in Appendix A of \inlinecite{Luoni11}. The upper half of the torus is progressively emerging without distortion. Our simple model does not take into account the Coriolis force, the effects of turbulence, and the deformations and reconnections occurring during the emergence. Its aim is only to provide a global description summarizing the typical observed evolution of the photospheric magnetic field during the emergence of an AR, more precisely the evolution of the magnetic tongues. It is also used to test the numerical method that we develop (\sect{pil}). The procedure consists of cutting the toroidal rope by successive horizontal planes, which play the role of the photosphere at successive times, and computing the magnetic-field projection in the direction normal to these planes. In this way, we obtain synthetic magnetograms with evolving tongues. The parameters that characterize the torus are: a main radius [$R$], a small radius [$a$], and the maximum axial-field strength [$B_{\rm max}$] (see Figure 12 of \opencite{Luoni11}). The torus center is located at a depth $-d$ below the photospheric plane. As the emergence proceeds, $-d$ varies between $-(R+a)$ and 0 (see Figure 2 of \opencite{Luoni11}). The sign and amount of magnetic twist of the modeled AR is given by imposing a uniform number of turns [$N_{\rm t}$] to the magnetic-field lines around the torus axis. $N_{\rm t}$ corresponds to half of the emerging torus. 

\fig{modN1} shows an example of a flux rope with $\Nt$ = 1 as it emerges at the photospheric level.  The flux-rope axis is along the $x$-direction.
The deformation of the polarities, the tongues, due to the presence of twist is evident in \fig{modN1}a. As the flux rope emerges, the polarities in the AR tend to separate and the tongues retract (\fig{modN1}b). In this simple model for a non-twisted flux-tube with $\Nt$ = 0 (not shown here) both polarities would have a round shape (symmetric in the $y$-direction) at all time steps, \ie\ tongues are not present.
In Figures~\ref{fig:modN1}a,b we plot black continuous lines that join the flux-weighted centers (that we will call barycenters) of the positive and negative magnetic polarities and the PIL, which has been computed as described below in \sect{pil}. The AR main bipolar axis joins the following to the leading barycenter; we define the bipole vector relative to this axis. As the flux rope emerges and magnetic tongues appear, the PIL is first inclined in the  East--West direction. When all of the flux has emerged and the polarities depart from each other, the PIL tends to be perpendicular to the main axis of the bipole. 

\begin{figure}[t]
\begin{center}
\includegraphics[width=.99\textwidth]{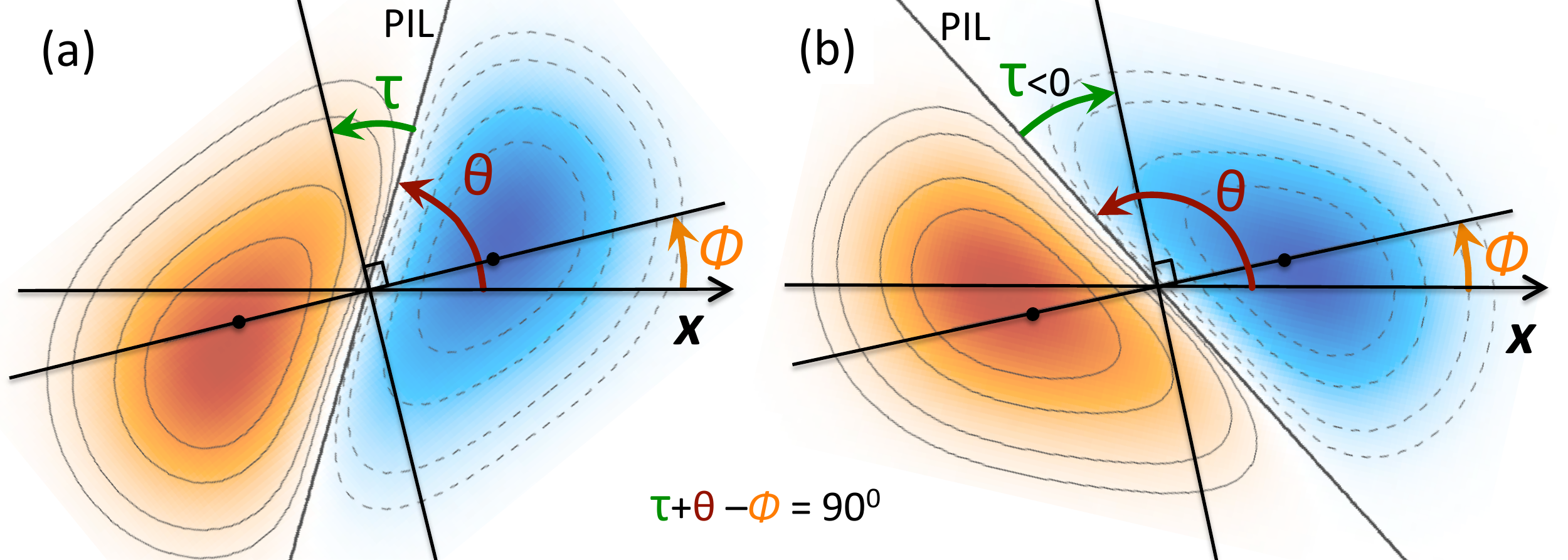}
\caption{The angles defined in \sect{aim} for a flux rope with (a) a positive and (b) negative magnetic helicity. All angles are positive when computed in a counter-clockwise direction.
$\phi$ is the tilt of the AR bipole and $\theta$ is the inclination of the PIL, both calculated from the East--West direction.  The tongue angle [$\tau$] is defined as the acute angle between the PIL direction and the direction orthogonal to the AR bipole.  With this definition, $\tau$ is a monotonically increasing function of the twist (\ie\ the number of turns) and it always has the same sign as the magnetic helicity of the modeled flux-rope.}
\label{fig:schema}
\end{center} 
\end{figure}

Next, we define the angles that are illustrated in Figures~\ref{fig:schema}a and \ref{fig:schema}b for a positive and negative twisted flux tube, respectively.
The angle [$\phi$] between the East--West direction and the bipole vector corresponds to the AR tilt.  
$\phi$ lies between $-90\degree$ and $90\degree$. 
The inclination of the PIL [$\theta$] is defined from the East--West direction. Since the PIL direction is defined modulo $180\degree$, we select  $\theta$ so that $\theta-\phi$ lies between $0\degree$ and $180\degree$.  Next, we define the tongue angle [$\tau$] as the acute angle between the PIL direction and the direction orthogonal to the AR bipole vector. $\tau$ is related to $\theta$ and $\phi$ as:
  \begin{equation}  \label{eq:tau}
   \tau = \phi - \theta + 90\degree  \,.
  \end{equation}
The above selected range for $\theta-\phi$ implies that $\tau$ lies between $-90\degree$ and $90\degree$, as required.  This definition of $\tau$ uses the bipole vector as a reference direction since it can be directly derived from magnetograms; however, as will be shown in 
\sect{Characteristics}, the bipole vector is only a first approximation to the orientation of the main axis of the bipole.

We now interpret $\tau$ in terms of the curved and twisted flux-tube model presented in \inlinecite{Luoni11}. 
When a flux-rope has no twist, the PIL is 
orthogonal to the bipole vector, implying $\tau=0$. When the flux-rope has positive twist, tongues are present as we show in \fig{schema}a. Then, $\tau$ is positive and it increases monotonically with the twist (due to an increasing contribution of the azimuthal component of the magnetic field to the component orthogonal to the photosphere).  When the flux-rope has negative twist, tongues are present as shown in \fig{schema}b; in this case,  $\tau$ takes a negative value that becomes monotonically more negative as the magnitude of the negative twist increases.

In Figures~\ref{fig:modN1}c,d we present the results derived from the model in the same way as we do later for the observed ARs. We plot the AR tilt, [$\phi$], the magnetic flux, and $\tau$ {\it versus} the time step. The plots show that there is a characteristic evolution of these AR parameters associated with the formation of the tongues. The presence of the magnetic flux in the tongues affects the computation of the polarity barycenters; that is, the estimated value of $\phi$. In our example, the tilt is initially around $-40\degree$; then, it monotonically goes to zero implying a large rotation of the bipole (\fig{modN1}c).
This rotation is only apparent because the flux-rope axis is in fact always in the $x$-direction. The presence of a finite twist also has a consequence for the magnetic-flux evolution, which reaches a maximum because of the added flux due to the projection of the azimuthal-field component on the direction orthogonal to the photosphere. As the emergence continues, the contribution of the azimuthal flux decreases and disappears when half of the torus has crossed the photospheric plane, \ie\ the flux-rope legs are orthogonal to this plane (\fig{modN1}c,d).  
In observed line-of-sight magnetograms such a decrease of the magnetic flux could be erroneously interpreted as flux cancellation. Furthermore, the sign of $\tau$ is well defined throughout the emergence; as shown in \fig{modN1}c, it tends to zero as the tongues retract and disappear.

Another parameter that characterizes the presence of magnetic tongues is the elongation [$E$] of the AR polarities.  \fig{modN1}{d} shows the evolution of the elongation of the polarities, computed as defined in Appendix~B of \inlinecite{Luoni11}. It is maximum at the beginning of the emergence phase and monotonically decreases as the flux-rope emerges. The signature of the tongues on the elongation weakens much earlier than that on $\tau$, as shown in \fig{modN1}d. However, the smaller $E$ is, the more uncertain is $\tau$, because $\tau$ is more easily affected by the presence of magnetic flux not belonging to the emerging flux-rope (\eg background flux in observed ARs).


\subsection{Computing the PIL Location}
\label{sec:pil}

In an observed magnetogram, the PIL can be a very complicated curved line. Therefore, the difficulty in tracing the PIL systematically makes it difficult to measure the $\tau$ and $\theta$ angles without ambiguity. However, only the global direction of the PIL is relevant to characterize the twist of the emerging flux-rope. Then, following \inlinecite{Luoni11}, we design a systematic procedure to compute a straight approximation to the PIL of an AR. In this latter article, a linear function [$a x + b y + c$] was fitted to the magnetogram [$B_o(x,y)$], where $a$, $b$, and $c$ are parameters to be determined by a least-squares fit and $x$ and $y$ are the spatial coordinates. This implies a strong contribution of the intense-field regions, which are not part of the AR tongues (since both functions are large, although different, away from the PIL). 
In particular, the distribution of the flux in the main AR polarities could have a significant effect on the computed PIL. Then, the fitting had to be spatially limited to the portion of the AR with tongues; this resulted in a cumbersome procedure in which the region where the fitting was done had to be selected by eye case by case.  We design below a new method to solve this difficulty.

  Our systematic procedure includes only the sign of the linear function $a x + b y + c$, avoiding large values away from the PIL.  Moreover, we use the absolute value of the observed magnetic field
in front of the fitted function. Then, we define the function $D(a,b,c)$ to be minimized as:

\begin{equation} 
\label{eq:def-dif}
D(a,b,c) = \int_{x,y} |\sign (a x + b y + c)~|B_o| - B_o|^2\; \rmd x \; \rmd y,
\end{equation} 

\noindent where $B_o = B_o(x,y)$ is the observed field, and the integration is done 
over the AR (more precisely, avoiding as much as possible surrounding fields not belonging to the AR). This integral approach is chosen in order to have robust results (\ie\ not sensitive to small polarities and noise of the magnetograms). The contribution of each magnetogram pixel to $D$ is chosen proportional to the square of the field strenght, so that low field pixels contribute less. 

The above procedure separates the magnetogram into a dominantly  
positive and a dominantly negative region. By construction the pixels of the magnetogram, where the field sign agrees with the sign corresponding to their side, do not contribute to $D$ because for those positions the function: $\sign (a x + b y + c)~|B_o| - B_o = 0$. Therefore, the best fit is decided by minimizing the contribution that lies in the ``wrong''-sign side of the PIL.  In particular, the strong- field regions of a bipolar AR are not contributing to $D(a,b,c)$ as they are on the correct-sign side of the PIL; in this way, the full AR can be included. More generally, the pixels far away from the computed PIL are either not contributing (correct sign) or have a constant contribution (wrong sign). Then, the minimization of $D(a,b,c)$ over the full magnetogram is in fact decided by a compromise among the pixels that are in the vicinity of the computed PIL.
This method has the advantage that it can be systematically applied to the full spatial extension of any bipolar AR.  The disadvantage is that the minimization of $D(a,b,c)$ does not reduce to a standard least-square fit as in \inlinecite{Luoni11}, and the method is numerically more expensive.

A more practical way to write the function $a x + b y + c$ is to use as parameters the position where the PIL crosses the segment that joins the barycenters of the main polarities and the PIL angle. If $\vec{r}_{\rm P}$ and $\vec{r}_{\rm F}$ are the vectors pointing to the barycenters of the positive and negative polarities, respectively, the point where the PIL intersects the vector $\vec{r}_{\rm P}-\vec{r}_{\rm F}$ is simply defined using a parameter [$s$], such that $0 \leq s \leq 1$ with $s = 0$ on $\vec{r}_{\rm F}$ and $s = 1$ on $\vec{r}_{\rm P}$. Then, the PIL intersection with the vector $(\vec{r}_{\rm P}-\vec{r}_{\rm F})$ is located at 

   \begin{equation} \label{eq:r(s)}
   \vec{r}(s) = (\xs ,\ys ) = \vec{r}_{\rm F}+s(\vec{r}_{\rm P}-\vec{r}_{\rm F}).
   \end{equation} 

\noindent The other parameter is the PIL angle, $\theta$. The function ($a x + b y + c$) is rewritten as $(x-\xs ) \sin \theta + (y-\ys ) \cos \theta$. Then, Equation (\ref{eq:def-dif}) turns out to be 

   \begin{equation} \label{eq:def-dif2}
   D(s,\theta) = \int_{x,y} \Big| \sign \big( (x-\xs ) \sin \theta + (y-\ys ) \cos \theta \big) ~|B_o| - B_o \Big|^2\; \rmd x \; \rmd y,
   \end{equation} 

\noindent which depends only on the parameters $s$ and $\theta$. Therefore, the aim is to find a global minimum for $D(s,\theta)$. This is done by discretizing the parameter space ($s_{n},\theta_{m}$) such that two nearby points, $s_n$ and $s_{n+1}$, are not farther than one-fourth of the magnetogram resolution and the step in $\theta_m$ is of about one degree. We compute Equation (\ref{eq:def-dif2}) on the resulting grid and find the global minimum value of $D(s_{n},\theta_{m})$.

\begin{figure}[t]
\begin{center}
\includegraphics[width=.99\textwidth]{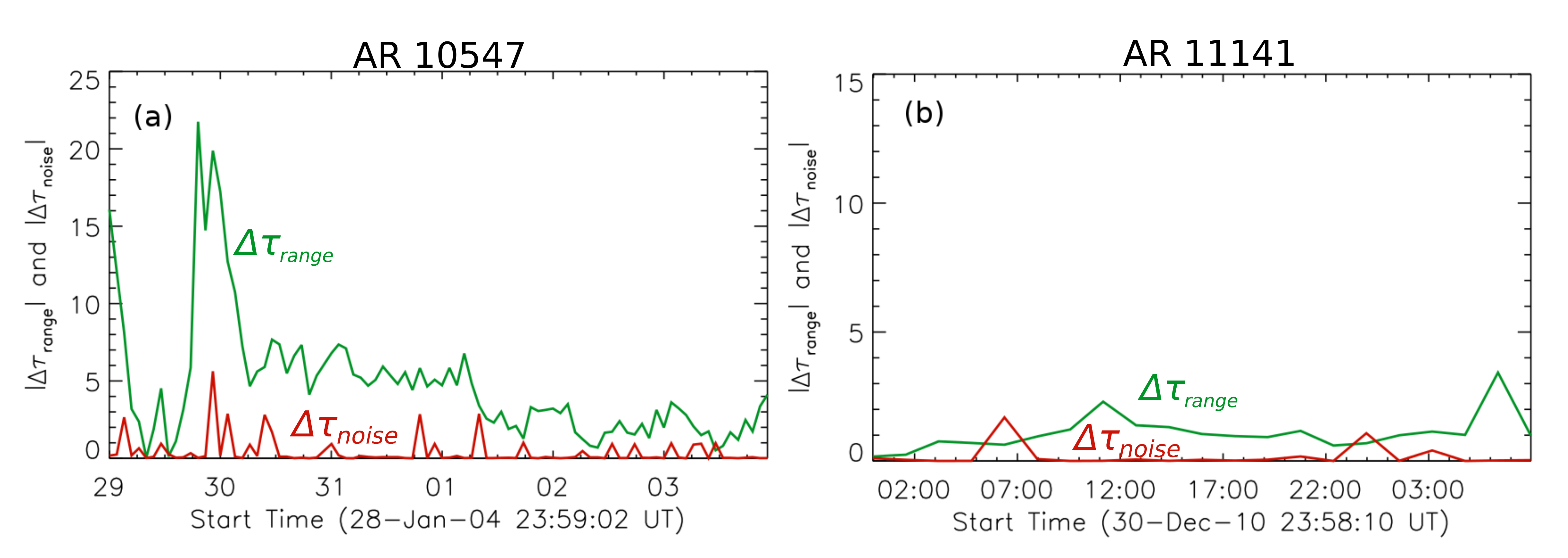}
\caption{Range of $\tau$, $\Delta \tau_{\rm range} = \tau_{\rm M}- \tau_{\rm m}$ (green continuous line), for the systematic errors due to wrong-sign side polarities, and $\Delta \tau_{\rm noise}$
(red continuous line), computed by adding random noise in the range [-10~G,10~G] to the magnetograms (see \sect{tau-error}). (a) AR 10547 observed by MDI (see \fig{10547}),  (b) AR 11141 observed by HMI (see \fig{11141}). These errors correspond to the emergence period when computed between the two vertical black dashed lines shown in \figs{10547}{11141}. 
}
\label{fig:error}
\end{center} 
\end{figure}


\begin{figure}[t]
\begin{center}
\includegraphics[width=.99\textwidth]{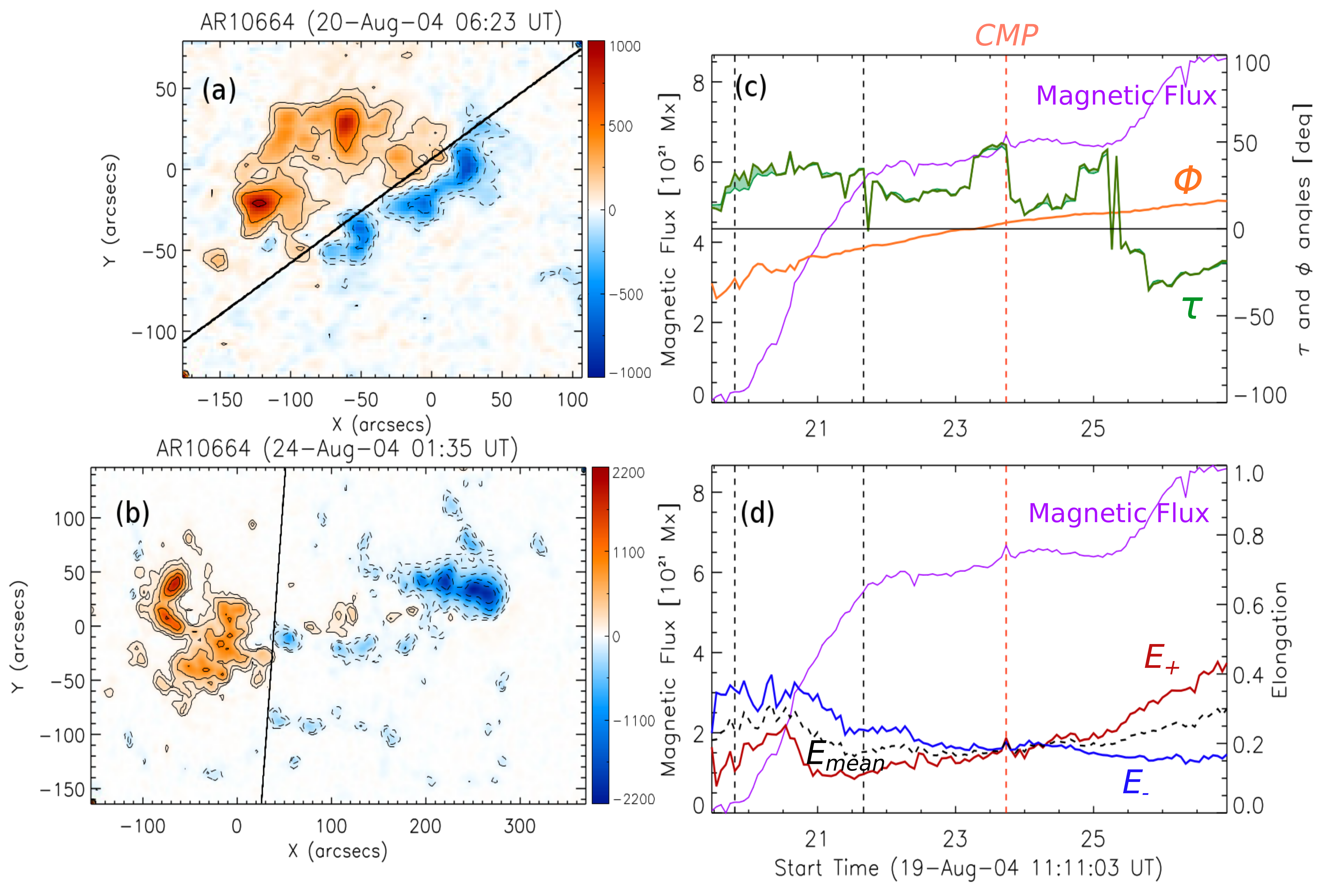}
\caption{Emergence of AR 10664 observed by MDI. Panels (a) and (b) correspond to magnetograms at two different times. The continuous (dashed) lines are isocontours of the positive
(negative) magnetic field component normal to the photosphere, drawn for 100, 200, 500, 1000, and 1500 G (and the corresponding negative values). The red and blue-shaded regions indicate positive and negative polarities, respectively. In these panels the horizontal (vertical) axis corresponds to the East--West (South--North) direction on the Sun.
(c) Evolution of the magnetic flux (violet continuous line), $\tau$ (green line), and the tilt angle (orange line) as defined in \fig{schema}. (d) Elongation of the positive (negative) polarity with red (blue) continuous line. The mean elongation is depicted with a black-dotted line. As a reference, we have added to this panel the evolution of the magnetic flux drawn in violet. The two vertical-dashed black lines mark the period of time in which we perform our computations and the orange vertical-dashed line indicates the AR central meridian passage (CMP).}
\label{fig:10664}
\end{center}
\end{figure}

\begin{figure}[t]
\begin{center}
\includegraphics[width=.99\textwidth]{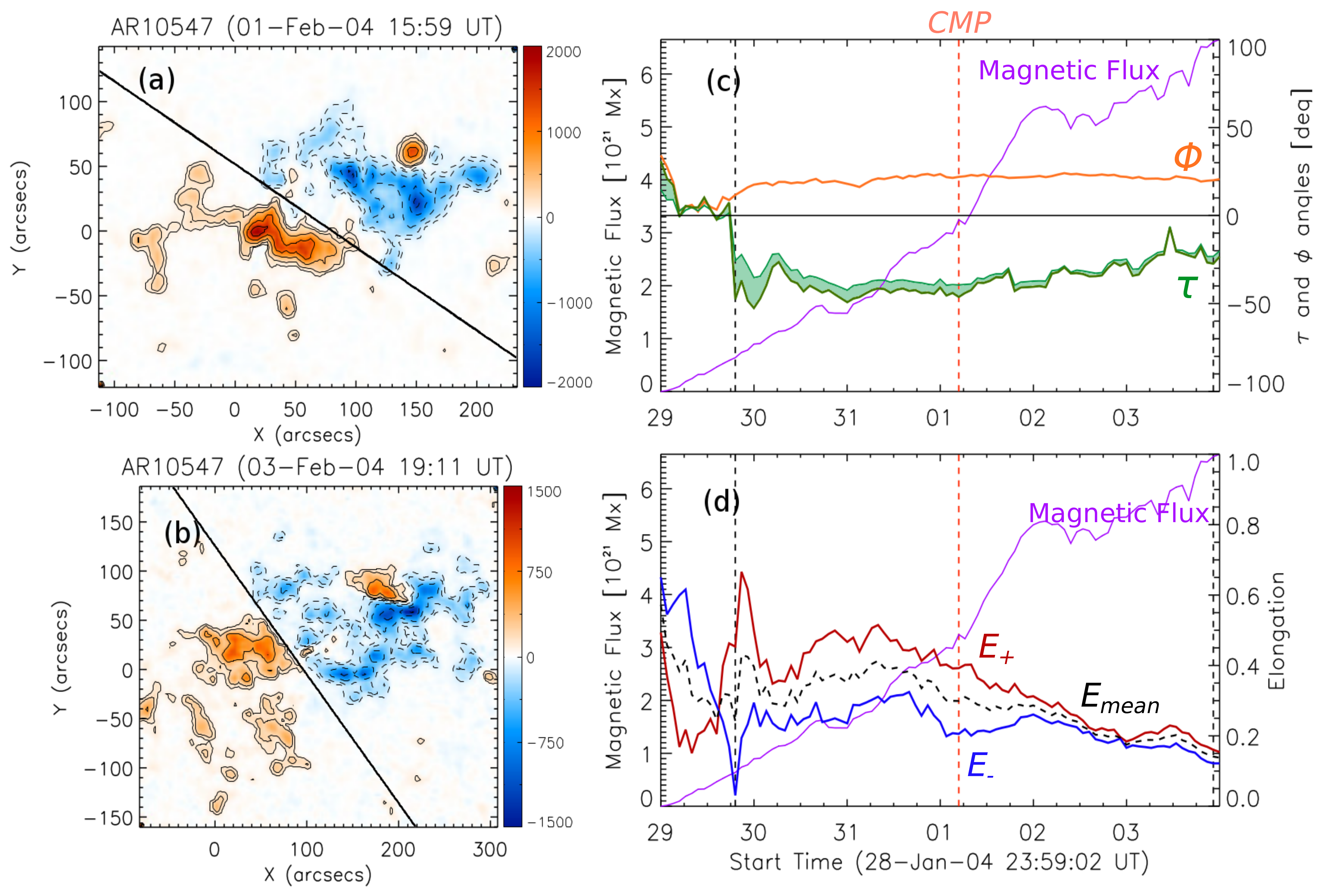}
\caption{Emergence of AR 10547 observed by MDI. Panels (a) and (b) correspond to magnetograms at two different times. The continuous (dashed) lines are isocontours of the positive
(negative) magnetic field component normal to the photosphere, drawn for 100, 200, 500, 1000, and 1500 G (and the corresponding negative values). The red and blue-shaded regions indicate positive and negative polarities, respectively. In these panels the horizontal (vertical) axis corresponds to the East--West (South--North) direction on the Sun.
(c) Evolution of the magnetic flux (violet continuous line), $\tau$ (green line), and the tilt angle (orange line) as defined in \fig{schema}. (d) Elongation of the positive (negative) polarity with red (blue) continuous line. The mean elongation is depicted with a black-dotted line. As a reference, we have added to this panel the evolution of the magnetic flux drawn in violet. The two vertical-dashed black lines mark the period of time in which we perform our computations and the orange vertical-dashed line indicates the AR central meridian passage (CMP). In this example, $\tau$ is stable and well determined (the systematic error range is narrow as shown by the green curve thickness).
}
\label{fig:10547}
\end{center}
\end{figure}

\begin{figure}[t]
\begin{center}
\includegraphics[width=.99\textwidth]{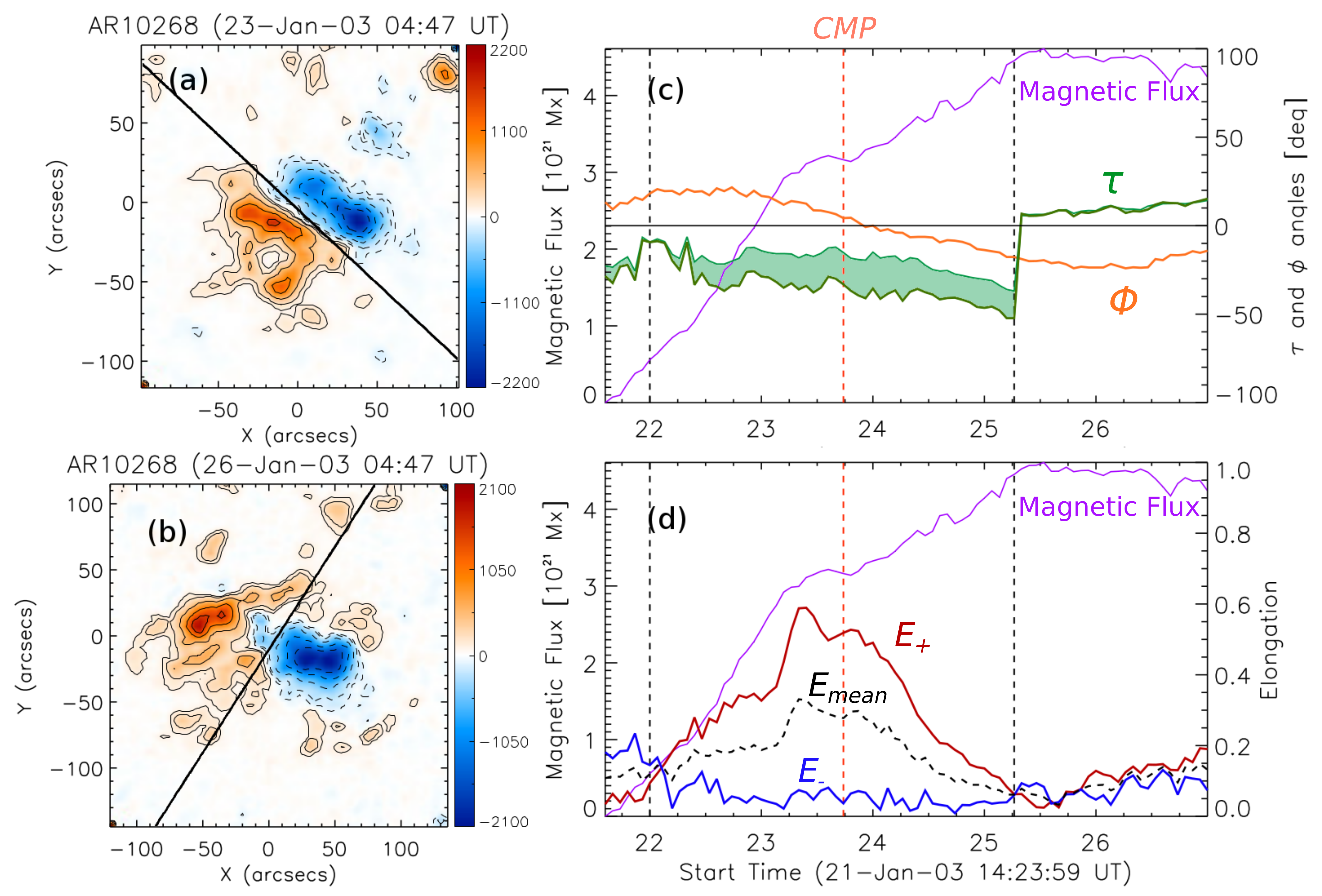}
\caption{Emergence of AR 10268 observed by MDI. Panels (a) and (b) correspond to magnetograms at two different times. The continuous (dashed) lines are isocontours of the positive
(negative) magnetic field component normal to the photosphere, drawn for 100, 200, 500, 1000, and 1500 G (and the corresponding negative values). The red and blue-shaded regions indicate positive and negative polarities, respectively. In these panels the horizontal (vertical) axis corresponds to the East--West (South--North) direction on the Sun.
(c) Evolution of the magnetic flux (violet continuous line), $\tau$ (green line), and the tilt angle (orange line) as defined in \fig{schema}. (d) Elongation of the positive (negative) polarity with red (blue) continuous line. The mean elongation is depicted with a black-dotted line. As a reference, we have added to this panel the evolution of the magnetic flux drawn in violet. The two vertical-dashed black lines mark the period of time in which we perform our computations and the orange vertical-dashed line indicates the AR central meridian passage (CMP). Same plots as in \fig{10664} for AR 10268, which has a magnetic flux lower by a factor of two. In this example, $\tau$ presents  
the largest uncertainties when compared to the ARs in \figss{10664}{10569} (as shown by the thickness of the $\tau$-curve).}
\label{fig:10268}
\end{center}
\end{figure}

\begin{figure}[t]
\begin{center}
\includegraphics[width=.99\textwidth]{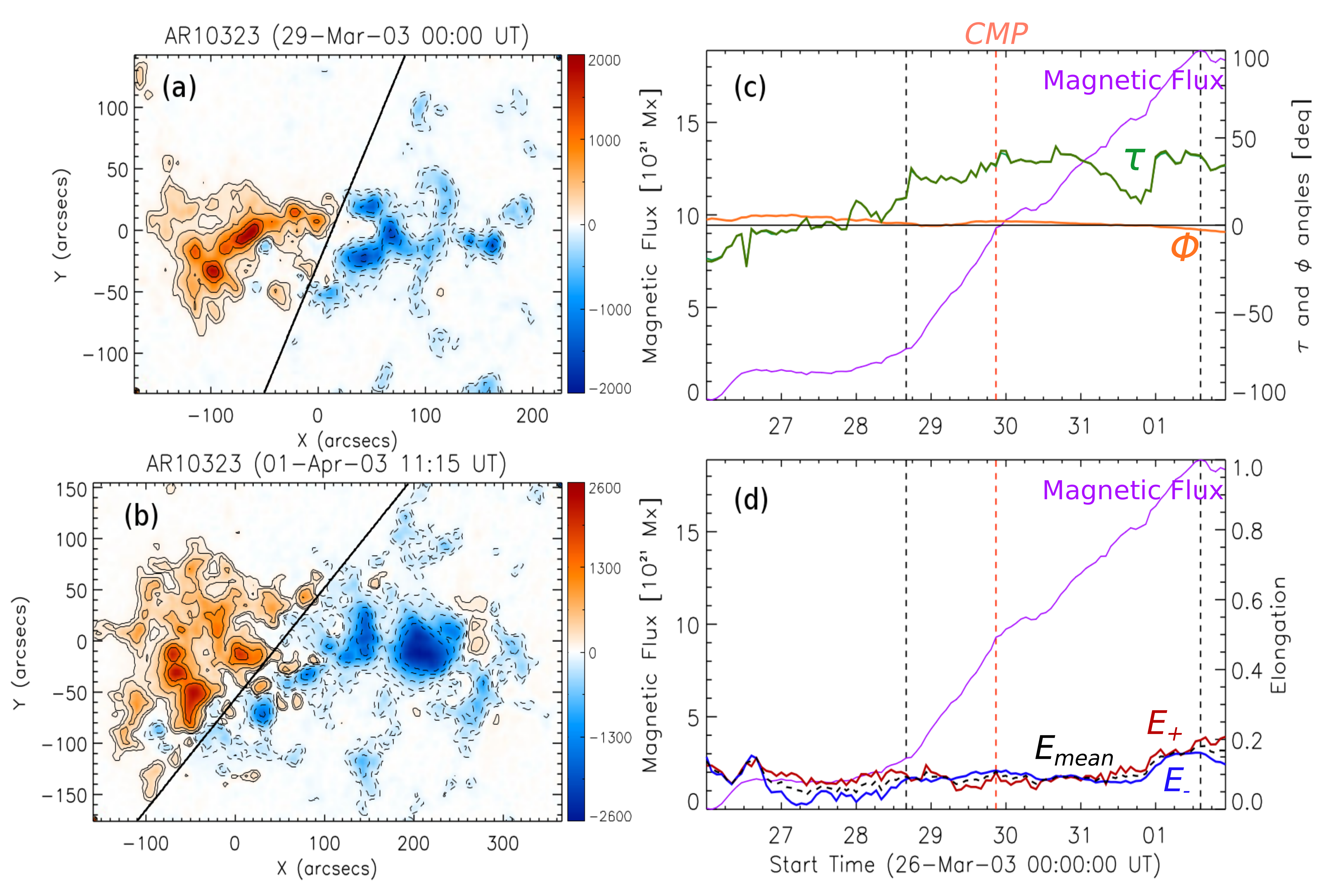}
\caption{Emergence of AR 10323 observed by MDI. Panels (a) and (b) correspond to magnetograms at two different times. The continuous (dashed) lines are isocontours of the positive
(negative) magnetic field component normal to the photosphere, drawn for 100, 200, 500, 1000, and 1500 G (and the corresponding negative values). The red and blue-shaded regions indicate positive and negative polarities, respectively. In these panels the horizontal (vertical) axis corresponds to the East--West (South--North) direction on the Sun.
(c) Evolution of the magnetic flux (violet continuous line), $\tau$ (green line), and the tilt angle (orange line) as defined in \fig{schema}. (d) Elongation of the positive (negative) polarity with red (blue) continuous line. The mean elongation is depicted with a black-dotted line. As a reference, we have added to this panel the evolution of the magnetic flux drawn in violet. The two vertical-dashed black lines mark the period of time in which we perform our computations and the orange vertical-dashed line indicates the AR central meridian passage (CMP). Similar plots as in \fig{10664} for AR 10323, which has a magnetic flux larger by a factor of two.}
\label{fig:10323}
\end{center}
\end{figure}

\begin{figure}[t]
\begin{center}
\includegraphics[width=.99\textwidth]{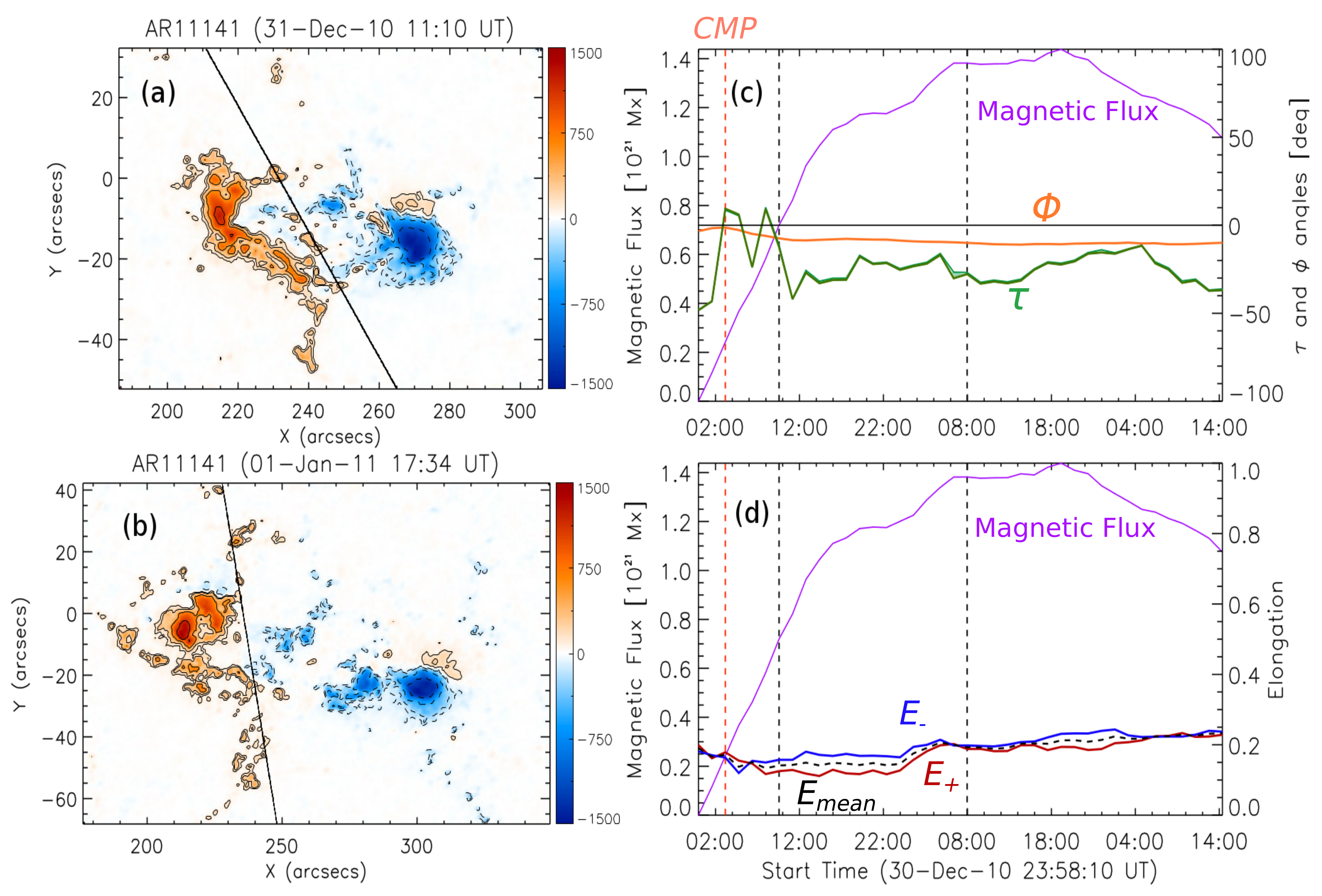}
\caption{Emergence of AR 11141 observed by HMI. Panels (a) and (b) correspond to magnetograms at two different times. The continuous (dashed) lines are isocontours of the positive
(negative) magnetic field component normal to the photosphere, drawn for 100, 200, 500, 1000, and 1500 G (and the corresponding negative values). The red and blue-shaded regions indicate positive and negative polarities, respectively. In these panels the horizontal (vertical) axis corresponds to the East--West (South--North) direction on the Sun.
(c) Evolution of the magnetic flux (violet continuous line), $\tau$ (green line), and the tilt angle (orange line) as defined in \fig{schema}. (d) Elongation of the positive (negative) polarity with red (blue) continuous line. The mean elongation is depicted with a black-dotted line. As a reference, we have added to this panel the evolution of the magnetic flux drawn in violet. The two vertical-dashed black lines mark the period of time in which we perform our computations and the orange vertical-dashed line indicates the AR central meridian passage (CMP). This AR has a very low magnetic flux, lower by a factor of six than the AR in \fig{10664}. }
\label{fig:11141}
\end{center}
\end{figure}

\begin{figure}[t]
\begin{center}
\includegraphics[width=.99\textwidth]{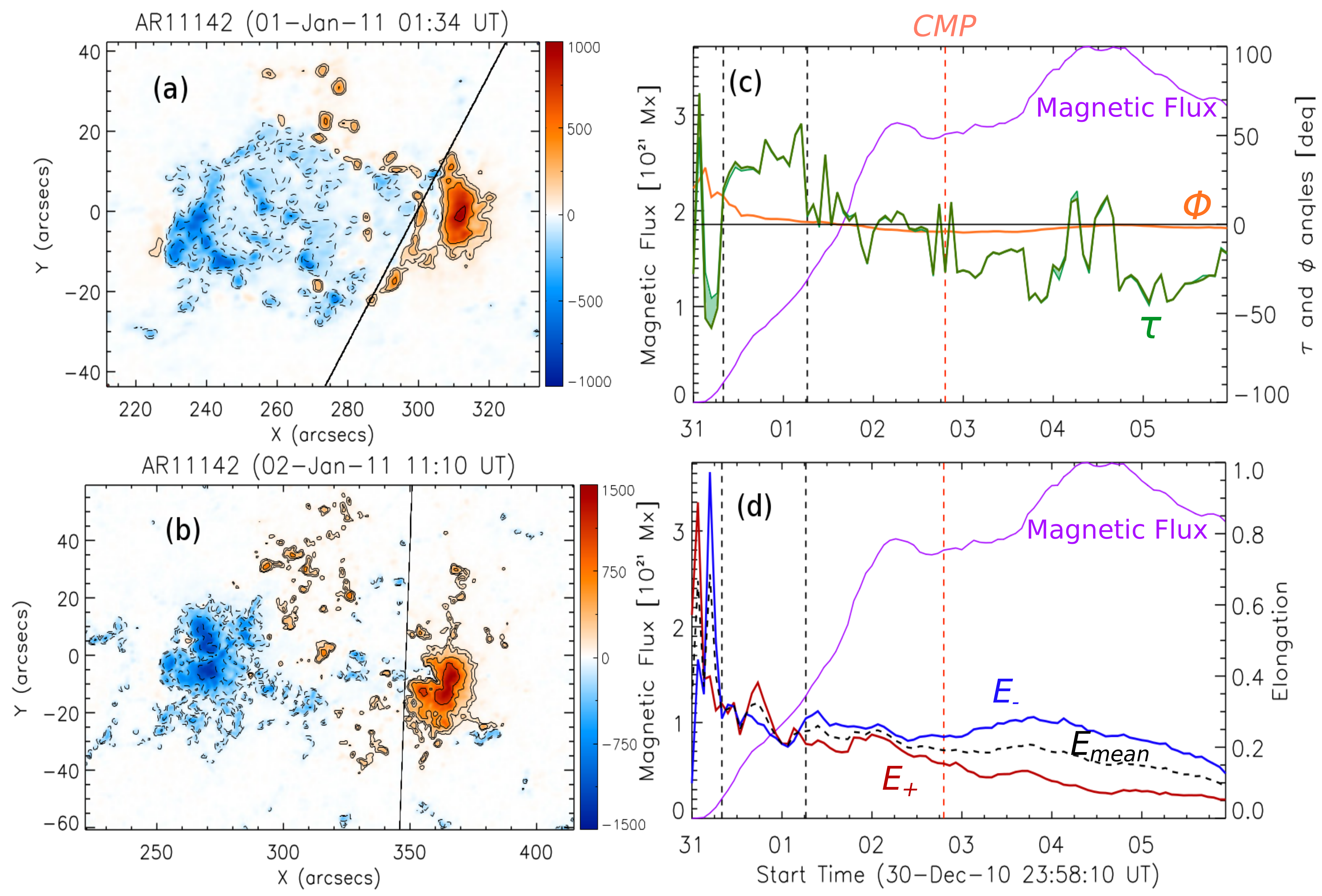}
\caption{Emergence of AR 11142 observed by HMI. Panels (a) and (b) correspond to magnetograms at two different times. The continuous (dashed) lines are isocontours of the positive
(negative) magnetic field component normal to the photosphere, drawn for 100, 200, 500, 1000, and 1500 G (and the corresponding negative values). The red and blue-shaded regions indicate positive and negative polarities, respectively. In these panels the horizontal (vertical) axis corresponds to the East--West (South--North) direction on the Sun.
(c) Evolution of the magnetic flux (violet continuous line), $\tau$ (green line), and the tilt angle (orange line) as defined in \fig{schema}. (d) Elongation of the positive (negative) polarity with red (blue) continuous line. The mean elongation is depicted with a black-dotted line. As a reference, we have added to this panel the evolution of the magnetic flux drawn in violet. The two vertical-dashed black lines mark the period of time in which we perform our computations and the orange vertical-dashed line indicates the AR central meridian passage (CMP). This AR has a magnetic flux lower by a factor of two than the AR in \fig{10664}.}
\label{fig:11142}
\end{center}
\end{figure}

\begin{figure}[t]
\begin{center}
\includegraphics[width=.99\textwidth]{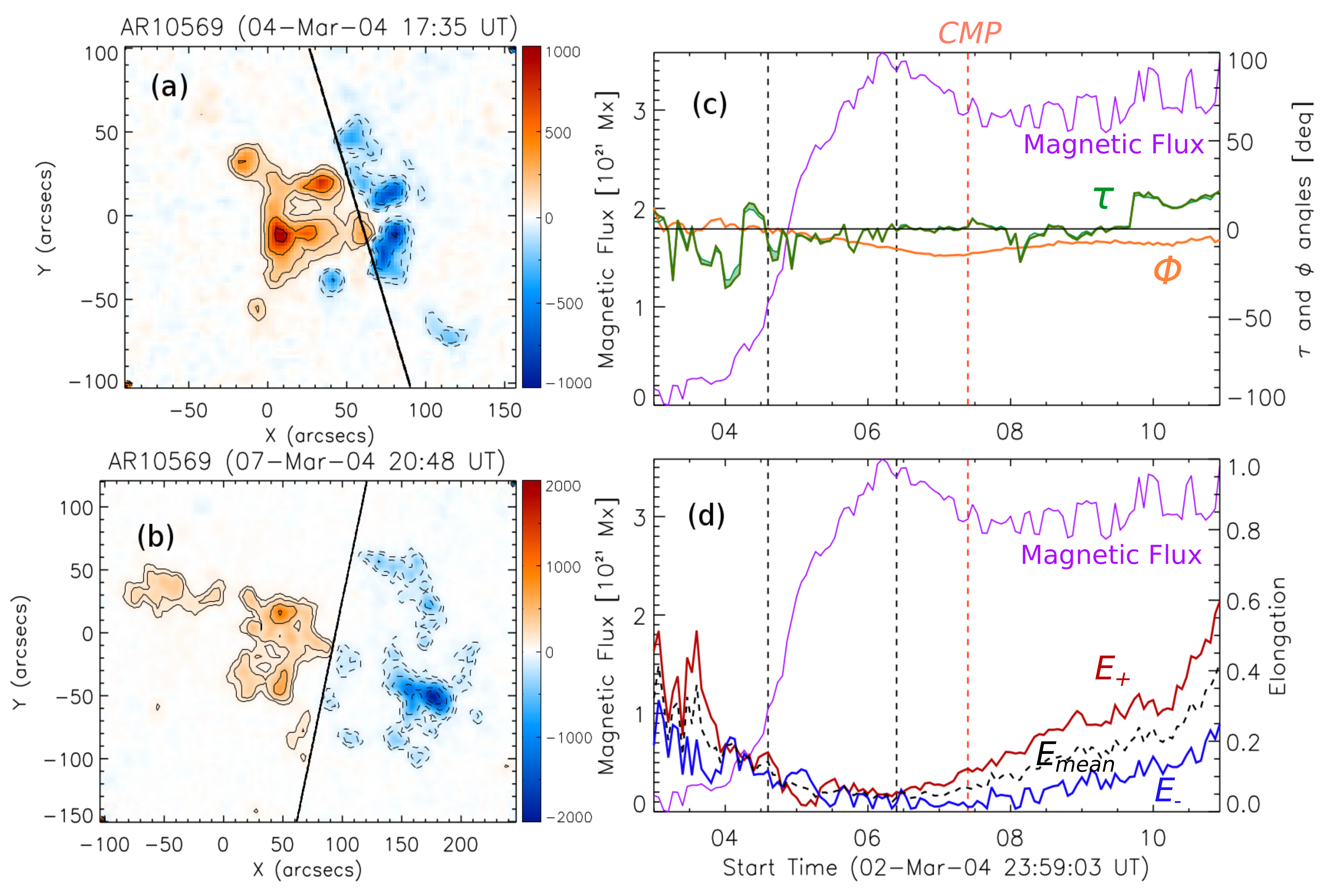}
\caption{Emergence of AR 10569 observed by MDI. Panels (a) and (b) correspond to magnetograms at two different times. The continuous (dashed) lines are isocontours of the positive
(negative) magnetic field component normal to the photosphere, drawn for 100, 200, 500, 1000, and 1500 G (and the corresponding negative values). The red and blue-shaded regions indicate positive and negative polarities, respectively. In these panels the horizontal (vertical) axis corresponds to the East--West (South--North) direction on the Sun.
(c) Evolution of the magnetic flux (violet continuous line), $\tau$ (green line), and the tilt angle (orange line) as defined in \fig{schema}. (d) Elongation of the positive (negative) polarity with red (blue) continuous line. The mean elongation is depicted with a black-dotted line. As a reference, we have added to this panel the evolution of the magnetic flux drawn in violet. The two vertical-dashed black lines mark the period of time in which we perform our computations and the orange vertical-dashed line indicates the AR central meridian passage (CMP). This AR has no significant magnetic tongues while its 
magnetic flux is comparable to that of the AR in \fig{11142}.
}
\label{fig:10569}
\end{center}
\end{figure}


\subsection{Errors in the PIL Inclination Angle and $\tau$ Angle}
\label{sec:tau-error}

When minimizing $D(s_{n},\theta_{m})$ in Equation~(\ref{eq:def-dif2}), we find some ARs in which disperse or complex polarities give multiple minima of $D$. From all of these minima, we select the global one. 

To estimate the error in the PIL inclination angle and, therefore, in the $\tau$-angle that we use to characterize the presence of  AR tongues, we evaluate how stable this global minimum is. To do so, we proceed as follows: First, we identify all of the pixels that are contributing to $D$ (\ie\ they are on the wrong-sign side of the PIL). Since those pixels are expected to be associated with parasitic polarities, not associated with the main polarities, we set the magnetic-field values of those pixels to zero and produce a modified magnetogram.
This new magnetogram has different characteristics compared to the original one, \ie\ the barycenters of the positive and negative main polarities are at a different location.  For this modified magnetogram, we compute a new bipole vector and PIL; then, we obtain a new set of parameters  [$s_1$ and $\theta_1$].  Furthermore, the number of pixels in the wrong-sign side for this new PIL will be different compared to the original magnetogram. This procedure of removing the pixels on the wrong-sign side is done iteratively until their total flux is low enough. More precisely, we stop the iteration when the magnetic flux in the pixels located in the wrong-sign side of the PIL is below 10~\% of the total unsigned flux of the AR.
We have fixed this percentage to save computation time and we have tested that taking it below 10~\% does not affect our results.  Moreover, if the change in $\tau$ obtained during a step is of the order of the step we have taken in $\theta$, we also stop the iteration.
Finally, following the previous procedure we determine the systematic error range in $\tau$, $[\tau_{\rm m}, \tau_{\rm M}]$,
within which all of the estimated $\tau$-values lie. The procedure provides a confidence interval defined by the presence or absence of polarities with the wrong sign on both sides of the PIL. Figures in \sect{results} illustrate the ranges in $\tau$ for several observed ARs.     
  
Another source of error in the determination of $\tau$ is the presence of noise in the measured magnetic 
field. We have added noise to the synthetic magnetograms generated from our flux-emergence model and also to the observed magnetograms. Considering that this noise is random, in the range [-10 G, 10 G] as for observations, and that we are working with an integral expression (Equation (4)), we expect the effect of this statistical error to be small or even negligible in the determination of the PIL direction. The error is given by the standard deviation of $\tau$ [$\Delta \tau_{\rm noise}$]. For the model magnetograms the error in $\tau$ stays, in general, below the angular step taken in $\theta$ ($0.9 \degree$). For the observed ARs, the maximum statistical error in $\tau$ is typically below 25~\% of the systematic error due to the presence of wrong-sign polarities on both sides of the computed PILs. \fig{error} shows the computed errors along the emergence of two ARs, one observed by MDI and one by HMI. The systematic error in $\tau$ is reported as the range within which its value can vary in \figss{10664}{10569}; this is shown by the green lines with variable thickness in these figures. All these results show that the tongues are well characterized by the angle $\tau$.

\section{The Evolution of the PIL in Observed ARs} 
      \label{sec:results}

\subsection{Generalities}
      \label{sec:Generalities}      

We describe the evolution of $\tau$ and other AR parameters, which can be affected by the presence of tongues, for some examples 
(\figss{10664}{10569}) in our set of emerging bipolar ARs (41) observed between early 2003 and early 2011.  These examples are selected to show the variety of cases present in the set. The minimum of the total magnetic flux in the first ten selected rectangles, before the time period shown for each AR, is subtracted from all the following ones, {\it i.e.} the magnetic-flux curve starts always from zero.

Several issues intrinsic to photospheric flux emergence can complicate our analysis and, thus, either limit the number of magnetograms used for an AR or produce fluctuating (though still coherent) results. We have dealt with these problems in most of the studied cases.

We are accompanying this article with movies (see the online material) displaying the evolution of these seven ARs. All of the movies show isocontours of the positive (negative) magnetic field in red (blue) continuous lines. The isocontour values are 50, 100, 500, and 1000~G (and the corresponding negative ones). Surrounded by a green dashed-line rectangle we depict the region selected to calculate all parameters in each frame. A black straight line joins the polarity barycenters (the extended violet line is included to guide the eye). The green straight line corresponds to the computed PIL. The black arch marks the angles between these two straight lines. The convention for the movie names is the AR NOAA number.

\subsection{Examples of Positive and Negative Twist}
      \label{sec:PositiveNegative}      

In \figs{10664}{10547} we show two examples of ARs with positive and negative magnetic helicity. AR 10664 (\fig{10664}) emerges in the southern hemisphere (S11) on 19 August 2004. Magnetic tongues are clear during a first  emergence episode (\fig{10664}a) until around 21 August at Noon. The magnetic flux reaches a maximum and starts increasing much slower. The value of $\tau$ starts fluctuating as the location of the PIL is affected by this slow magnetic flux increase in between the two main AR polarities. By early 24 August the AR PIL is almost orthogonal to the East--West direction (\fig{10664}b). After 25 August, the magnetic flux in this secondary flux emergence increases more steeply but tongues are not clearly discernible. The sign of $\tau$ during the first episode corresponds to a positively twisted flux rope (\fig{10664}c). Both the evolution of $\tau$ and $\phi$ follow the one shown in \fig{modN1} during the period marked between the two vertical black dashed lines, {\it i.e.} the distance between the curves of $\tau$ and $\phi$ remains almost constant. In \sect{Characteristics} (see Equation~(\ref{eq:tauc})), we will show that this distance is related to a corrected value of $\tau$.

AR 10547 (\fig{10547}) is also a southern hemisphere (S9) region that appears on 29 January 2004. This AR emerges quite slowly, \ie\ AR 10547 reaches a magnetic flux of 
$7 \times 10^{21}$ Mx after six days of emergence. From late 30 January until early 3 February, we observe clear magnetic tongues with the typical shape corresponding to a negative twist (\fig{10547}a,b).  The distance between $\tau$ and $\phi$ remains approximately constant and slowly decreases as the AR has fully emerged and tongues disappear (\fig{10547}c), in a similar way to the flux-rope model (mirroring \fig{modN1} to have the negative twist case).  However, $\phi$ remains approximately constant and this implies that the presence of tongues is not affecting the direction of the AR tilt, {\it i.e.} there is no apparent rotation of the AR main bipole because of the presence of tongues (see \sect{Characteristics}). In this particular case, the analysis is not continued until reaching the maximum flux value because the AR is farther from disk center than our imposed selection criterion.

Since the polarities are not equally elongated, as happens in the modeled ARs, \fig{10664}d and \fig{10547}d show the elongation of the positive [$E_{+}$] and negative [$E_{-}$] polarities and their average [$E_{\rm mean}$] independently. The definition of these parameters is the same as that of Equation (14) in \inlinecite{Luoni11} and $E_{\rm mean}$ is the standard average between $E_{+}$ and $E_{-}$.
Regarding the elongation of the polarities, in both cases its evolution is consistent with that of the model AR (\fig{modN1} and its mirror for a negative twist flux rope) between the two vertical black-dashed lines, which mark the period of time that we analyze (as in \figsss{10664}{c}{10547}{c}). After some strong fluctuations of the elongation due to the difficulty in measuring the parameters at the emergence initiation (in particular for AR 10547), in both cases, as in most of the studied ARs, there is a clear asymmetry between the elongations of the polarities. This is related to the known observed asymmetry of the polarities in bipolar ARs, {\it i.e.} the magnetic flux in the preceding polarity is concentrated, while it is dispersed in the following one (\opencite{vanDriel90}; \opencite{Fan93}; \opencite{vanDriel14}). As discussed in \sect{aim}, the flux-rope model used in this article does not take into account these asymmetries. However, it is noteworthy that our results show that these asymmetries do not affect the applicability of the method to determine the direction of the PIL. The general trend for the elongation curves is a marked decrease as the AR reaches maximum flux for each emergence, while $\tau$ is still measurable.
  
\subsection{Examples of Low and High Magnetic Flux}
      \label{sec:LowHighFlux}      

In \figs{10268}{10323} we show two other examples. These have a maximum magnetic flux lower by a factor of $\approx 2$, for AR 10268, and higher by a factor of $\approx 2$, for AR 10323, when compared to the previous cases. 

AR 10268 is a northern-hemisphere region that starts emerging on 21 January 2003 (\fig{10268}c). In fact two coherent emergences (maintaining a similar $\tau$-value) are observed in the period marked by the two vertical black-dashed lines in \fig{10268}c. By around 9:35 UT on 25 January, positive flux emergence at the North of the AR, accompanied by a less remarkable negative emergence nearby, makes the value of $\tau$ jump to low positive values as shown in the accompanying movie, {\it i.e.} the inclination of the PIL changes drastically guided by these emergences (\fig{10268}b). Apart from this, the evolution of $\tau$ and $\phi$ follow closely the one expected in the mirror of \fig{modN1}; the distance between these two curves remains approximately constant as for AR 10547 (\fig{10547}c). 

AR 10323 is a southern hemisphere AR (S07) that presents a first emergence around 26 March, tongues are not clear at this stage and $\tau$ fluctuates around zero. By 28 March at around 12:00 UT a second emergence with clear positive twist is observed (\fig{10323}c). The value of $\tau$ starts increasing and remains almost constant, except for a decrease between 31 March around 06:00 UT and the beginning of 1 April. During that period the emergence of positive and negative flux at the South of the positive polarity 
decreases the PIL inclination. Later as the new negative polarity extends and merges with the main one, the value of $\tau$ increases and goes back to the previous value. During this emergence period, $\phi$ remains constant indicating that the magnetic flux in the tongues is low and does not affect the AR tilt (\fig{10323}c). 

The evolution of the elongation of the tongues in AR 10268 is affected by the two emergences, but finally it decreases as expected in the AR flux emergence model (\fig{10268}d). For AR 10323 the elongation of the tongues remains approximately constant for the second flux emergence (\fig{10323}c). Summarizing, in these two examples the amount of magnetic flux has no significant effect on the detection of the tongues when using the evolution of $\tau$ in contrast to the elongation $E$.

\subsection{Active Regions Observed by HMI}
      \label{sec:ARs_HMI}      

The next examples (\figs{11141}{11142}) correspond to two small bipolar ARs observed in late 2010 with HMI, when it was fully operational. In order to compare the results derived from both instruments and the influence of the instrument spatial resolution on our results, we have, on one hand, reduced the HMI cadence by taking one magnetogram every 96 minutes but keeping its spatial resolution and, on the other hand, reduced HMI spatial resolution to match that of MDI. For these two experiments, the data are processed in the same way as the MDI magnetograms. 

In \figsss{11141}{a,b}{11142}{a,b} we show the results of the first experiment. It is difficult to see a long clear evolution of the magnetic tongues in both ARs because of their very low magnetic flux, which is contaminated by the background noise. The latter is more remarkable for AR 11142, in which $\tau$ shows strong fluctuations. However, it has a clear predominant sign, negative for AR 11141 and positive for AR 11142, which is evident during the main emergence phase (\figsss{11141}{c}{11142}{c}). For both ARs, the value of $\phi$ remains constant indicating that the flux in the tongues is low enough to not affect the direction of the AR bipole vector. The elongation of the tongues is approximately constant for AR 11141, while it decreases in time for AR 11142.  In this later case the preceding polarity has a lower elongation compared to the following one.

In summary, despite the increase in the spatial resolution (HMI images have four times higher resolution than those of MDI), the results related to the presence of tongues are coherent with those of the previous examples. Finally, when decreasing the spatial resolution of HMI to MDI (results not shown), we found no significant change concerning the direction of the PIL and its evolution. The difference between the mean value of $\tau$ using HMI spatial resolution and reducing it to MDI spatial resolution stays below 2$\degree$; the same happens for the maximum value of $\tau$.
 
\subsection{Example of an AR Without Significant Tongues}
      \label{sec:NoTongue}      

As a last example, Figures~\ref{fig:10569}a,b illustrate the emergence of AR 10569. This is a southern hemisphere (S11) region that appears in March 2004. The magnetic flux in the AR reaches its maximum in two days. During that period the sign of $\tau$ is not well defined, fluctuating around zero (\fig{10569}c), while $\phi$ remains almost constant. However, the elongation of the polarities suggests that tongues could be present (\fig{10569}d), although not important. Examining some individual magnetograms on late 3 March and the beginning of 4 March, we are able to identify magnetic tongues that correspond to a negative twist. This AR is one of the few cases in which the variation of $\tau$ is not sufficient to determine the sign of the twist and visual inspection is needed to identify the presence of tongues and their shape.


\section{Active Region Properties Derived from Magnetic Tongues} 
\label{sec:Global} 

\subsection{General Definitions}
      \label{sec:Definition}      

To be able to compare observed ARs with different characteristics and reach some general conclusions, we will use the average and maximum values of the computed parameters. For instance, we define $\tauMean$ as the mean of $\tau$ calculated over a limited time period during the emergence, when the magnetic flux in the tongues is dominating the background flux and the mixed (sea-serpent) polarities 
(see \sect{results}). We present below a histogram of its absolute value [$|\tauMean|$]. We also define $|\tauMax|$ as the maximum value of $|\tau|$ for the same period of time. Similarly, we introduce the mean tilt angle [$\phiMean$] and the absolute value of the maximum difference between $\phi$ and $\phi_{\rm max~flux}$, where the later is the bipole tilt at maximum magnetic flux. The value $|\phi -\phi_{\rm max~flux}|_{\rm max}$ represents the maximum departure from the estimated flux-rope axis. In the same way, we introduce the average and maximum absolute values of the number of turns in half the emerging toroidal flux rope model along the studied time period. After correction because of the influence of the AR tilt angle, these signed numbers will be the parameters that we use as a measure of the observed AR twist (see \sect{Characteristics}). The results for the 41 ARs are listed in Table~\ref{tab:ars} of the Appendix.

All of the values defined in the previous paragraph are computed only in the temporal ranges corresponding to the main flux emergence, as marked between the vertical-dashed lines in \figss{10664}{10569}. 
The criteria to select these intervals are the following: First, we identify the main phase of the emergence as the temporal interval during which the magnetic flux has a pronounced increase. Then, by direct inspection of the magnetograms we refine the chosen interval by minimizing the effects of the background and of secondary emergences in the selected rectangles. In some cases, the background has an important effect at the beginning of the AR emergence; thus, affecting the determination of the PIL inclination. For these cases we remove the first magnetograms until the AR polarities have gained enough strength to make the effect of the background negligible. Furthermore, during flux emergence, the known sea-serpent effect (\opencite{Strous96}; \opencite{Pariat04}) contaminates the determination of the PIL as it creates chains of minor polarities. In cases in which secondary emergences are observed (\eg\ \fig{10664}), the structure of the ARs becomes increasingly complex making the PIL determination unreliable. In those cases we include in the analysis only the main phase of the first emergence. This is done for the 41 analyzed ARs.

\begin{figure}[t]
\begin{center}
\includegraphics[width=\textwidth]{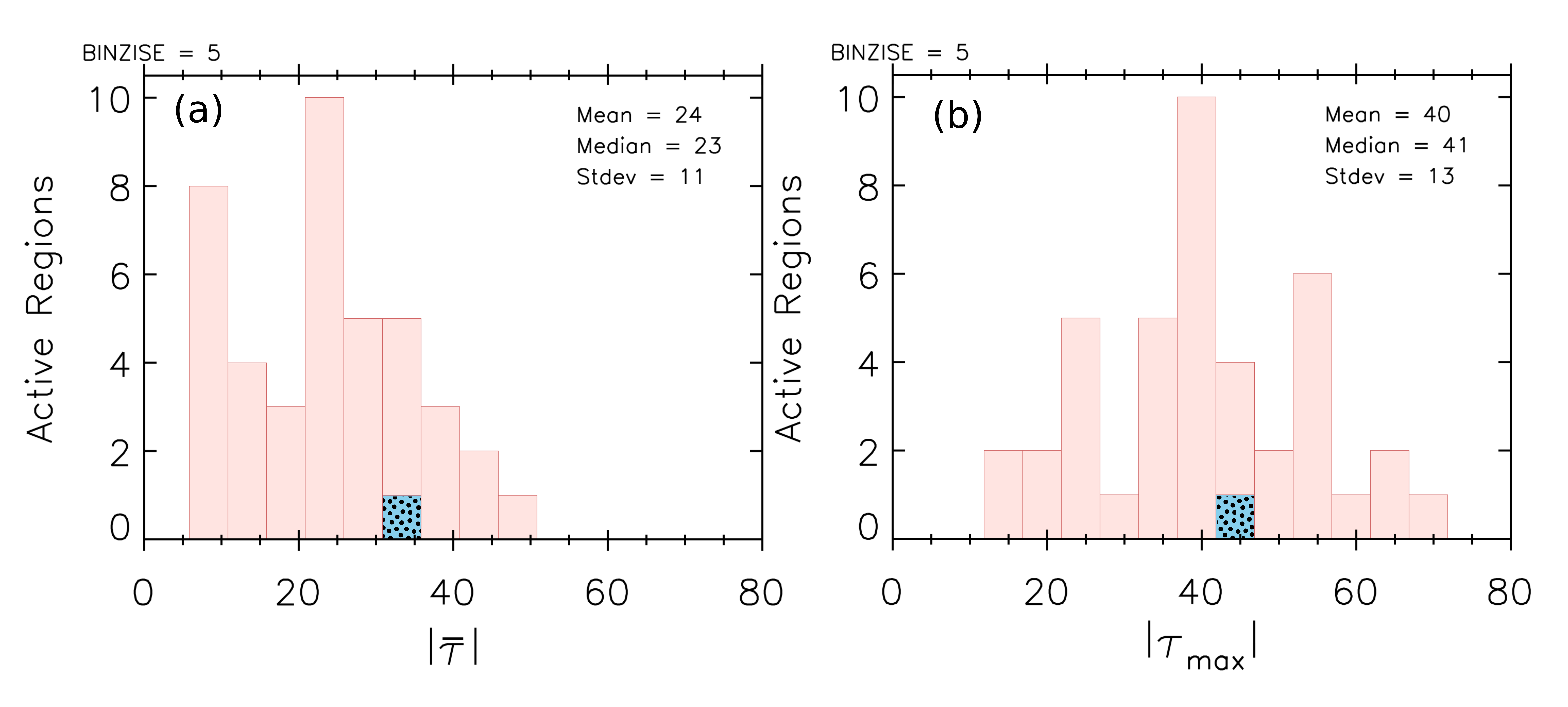} 
\caption{Histograms of the absolute values of (a) the mean of $\tau$, $|\tauMean|$, and (b) the maximum of $\tau$, $|\tauMax|$, for all of the studied ARs. The values of the mean, median, and standard deviation are shown as insets in both panels. The values of $|\tauMean|$ and $|\tauMax|$ for the AR-emergence model in \fig{modN1} ($\Nt =1$) are shown with a light blue rectangle with black-dotted patern.}
\label{fig:histo-tau}
\end{center}
\end{figure}

\begin{figure}[t]
\begin{center}
\includegraphics[width=\textwidth]{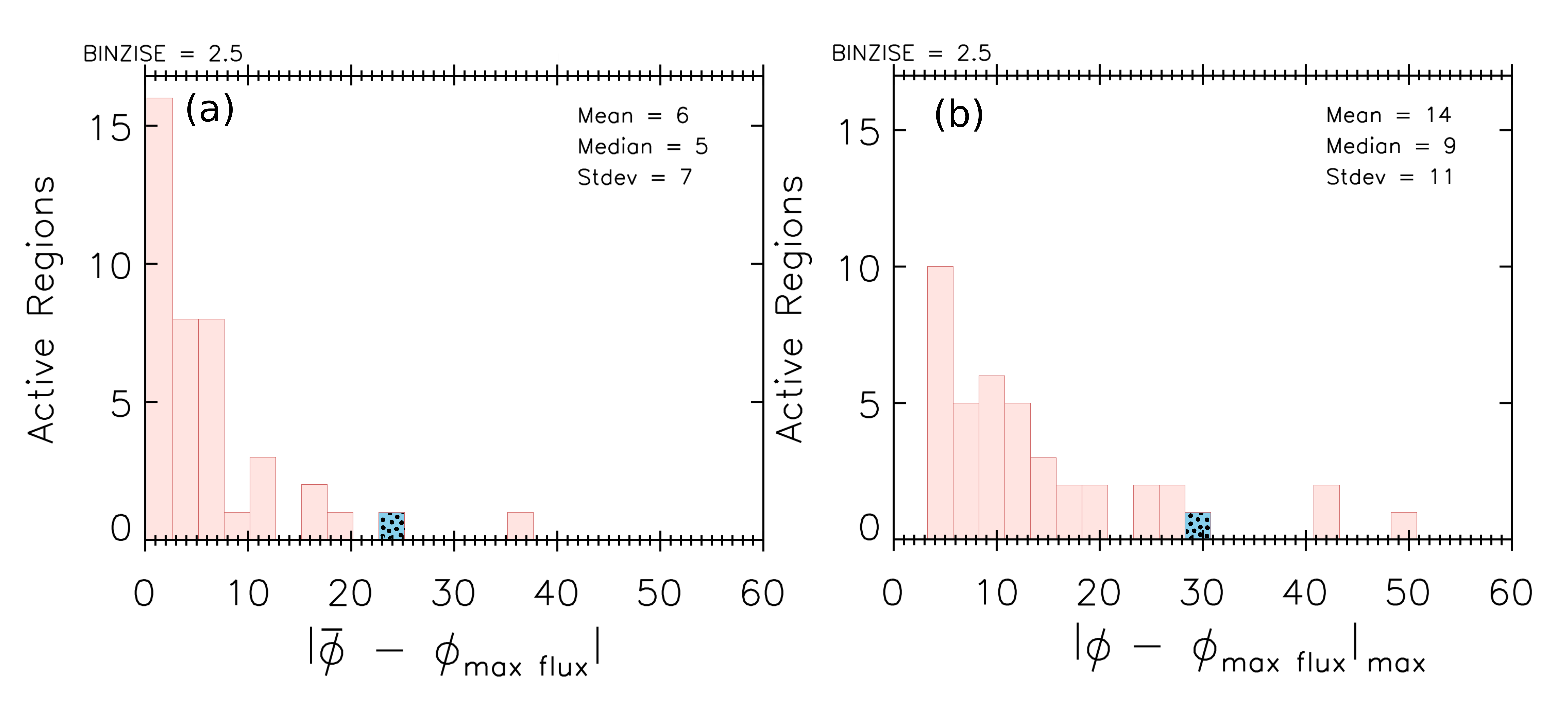}
\caption{Influence of the tongues on the bipole rotation during the emerging phase. Histograms of (a) $|\phiMean -\phi_{\rm max~flux}|$ and (b) $|\phi -\phi_{\rm max~flux}|_{\rm max}$, as defined in the text. The light blue rectangle with black-dotted pattern corresponds to our AR emergence model shown in \fig{modN1} ($\Nt =1$). 
}
\label{fig:histo-phi}
\end{center}
\end{figure}

\subsection{Tongue Characteristics}
      \label{sec:Characteristics}      

In \fig{histo-tau} we show histograms of the absolute values of the mean and maximum $\tau$  [$|\tauMean|$ and $|\tauMax|$] for the 41 studied ARs.
$|\tauMean|$ (\fig{histo-tau}a) is approximately 24$\degree$ in close correspondence with the median of the distribution.
From \fig{histo-tau}b, $|\tauMax|$ has a mean value of 40$\degree$, with a median of 41$\degree$. Although the distribution of $|\tauMax|$ is slightly more spread because it depends on a single measurement for each AR, it looks more Gaussian than the distribution of $|\tauMean|$. The values of $|\tauMean|$ and $|\tauMax|$ for the AR emergence model (light blue rectangles in Figures~\ref{fig:histo-tau}a,b), shown in \fig{modN1}, are slightly above the mean/median values for the observed ARs. This is a first indication that this simple model represents well the typical behavior of $\tau$ for ARs during the emergence phase.

The way in which the magnetic flux in the tongues can influence the variation of the AR tilt angle can be understood  from \fig{histo-phi}, where we plot the histograms of $|\phiMean -\phi_{\rm max~flux}|$ and 
$|\phi -\phi_{\rm max~flux}|_{\rm max}$. We recall that $\phi_{\rm max~flux}$ corresponds to the tilt angle when the AR has reached its maximum flux, {\it i.e.} in general when the bipole has fully emerged. It is clear from both histograms that for the majority of the cases studied, the magnetic flux in the tongues has a small influence on the AR tilt. This is evident from the mean values of the histograms (6$\degree$ in \fig{histo-phi}a and 14$\degree$ in  \fig{histo-phi}b). However, a few active regions present values in the tails of the distributions. Moreover, although the values of $|\tauMean|$ and $|\tauMax|$ for our emergence model stay close to the mean values for the observations, the magnetic flux in the tongues is much higher than in our observed cases; the light-blue rectangles with black-dotted pattern in both histograms of \fig{histo-phi} are located at large values (this implies that the tongues affect the tilt in the model).

\begin{figure}[t]
\begin{center}
\includegraphics[width=\textwidth]{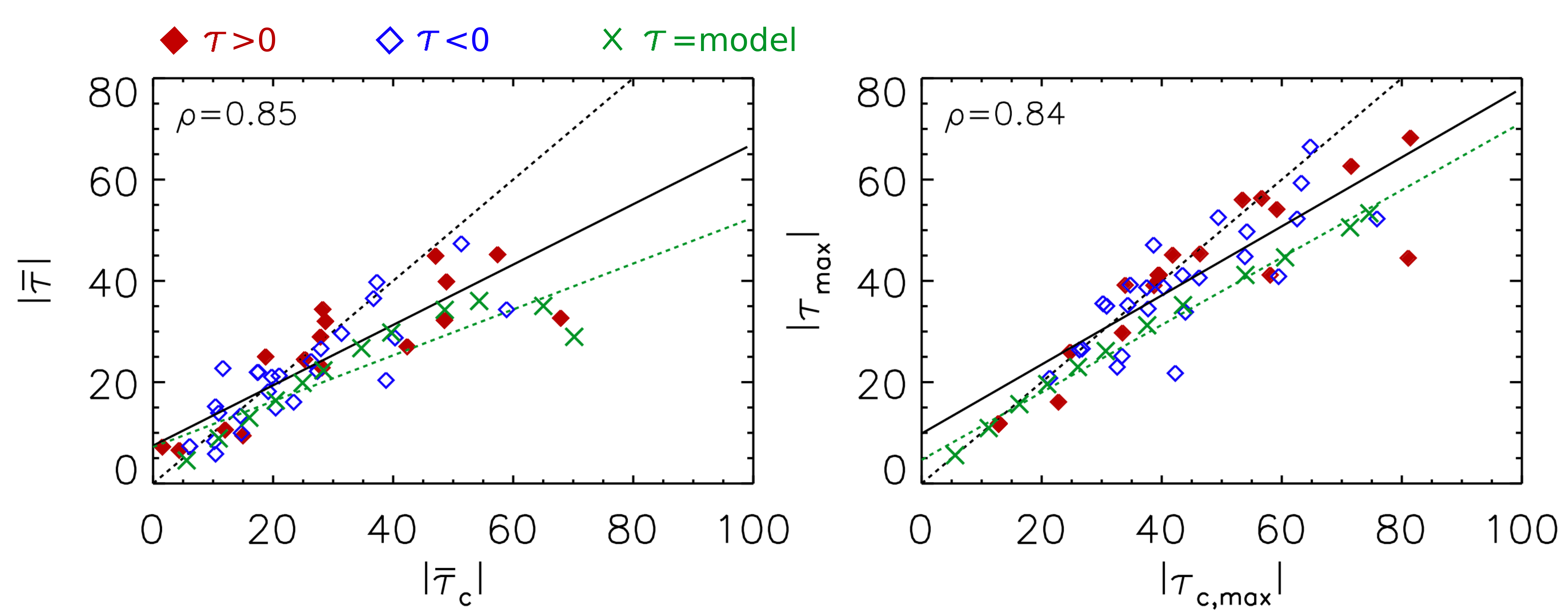}
\caption{Correlation of (a) $|\taucMean|$ with $|\tauMean|$ and (b) $|\taucMax|$ with $|\tauMax|$ for all of the studied ARs. The black-dotted lines mark the equality of abscissa and ordinate. The black-continuous lines are the least-squares fits with a straight line to all of the data points, blue-open (red-filled) diamonds correspond to $\tauMean<0$ ($\tauMean>0$).  The results from the analytical model (see the case $\Nt =1$ in \fig{modN1}) are plotted with green crosses for $\Nt =0.05, 0.1, 0.15, 0.2, 0.25, 0.3, 0.4, 0.5, 0.75, 1, 2, 4 $. 
The green-dotted lines are the least-squares fits with a straight line to all of the model points. These show a global tendency comparable to the observations for a fraction of the ARs.}
\label{fig:correl-tau}
\end{center}
\end{figure}

   Since the magnetic tongues have an effect on the bipole tilt for some ARs, we define a corrected tilt angle. We assume that the flux rope is not rotating significantly during emergence, so that its axis direction can be inferred from the tilt angle when the tongues have significantly retracted and before another emergence or dispersion further modifies the bipole tilt. We select the time of maximum flux as that satisfying both conditions. Then, we define the corrected angle [$\tauc$] as
  \BE \label{eq:tauc}
  \tauc (t) = \tau (t) - \phi (t) + \phi_{\rm max~flux} \, ,
  \EE
  and we compute an average [$\taucMean$] and a maximum value [$|\taucMax|$] as above for $\tau$.  An important fraction of the ARs are found in the vicinity of the lines $\taucMean=\tauMean$ and $|\taucMax|=|\tauMax|$ (black-dotted lines in \fig{correl-tau}). The dispersion around these lines is interpreted as an intrinsic rotation of the bipoles (\eg\ due to the buffeting of the convective cells).   However, a fraction of ARs have significantly larger values of $\taucMean$ than $\tauMean$; the same happens for the maximum values.  Indeed, the least-squares fit of all of the data points with a straight line is found close to the results of our emerging model by varying the number of turns [$\Nt$] (black continuous line and green crosses in \fig{correl-tau}). This correspondence is closer for a fraction of the ARs. Again, this shows that this simple model describes well typical ARs in terms of the PIL orientation during flux emergence. A broad range of $\Nt$ values is needed to interpret the analyzed ARs.
   
\begin{figure}[t]
\begin{center}
\includegraphics[width=\textwidth]{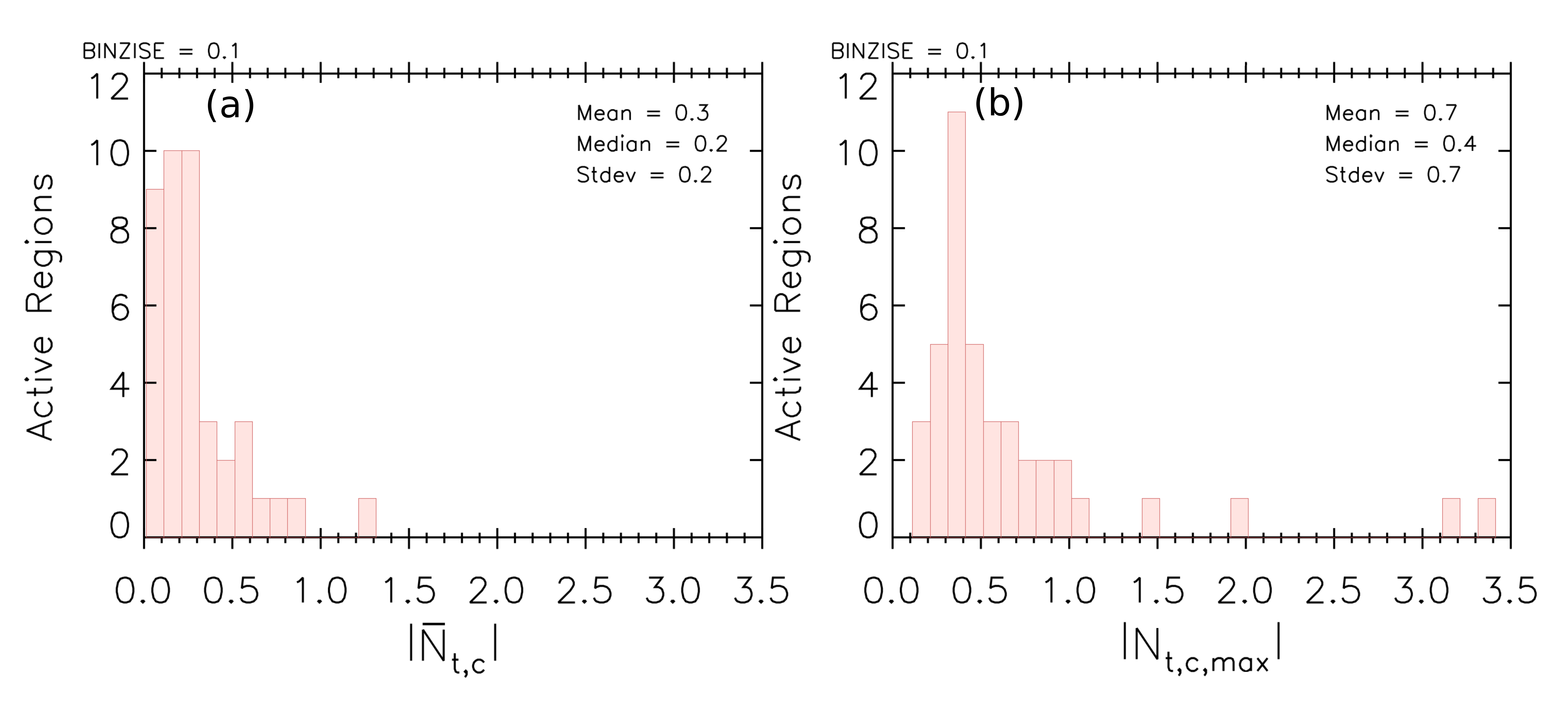}
\caption{Histograms of $|\NtcMean|$ and $|\NtcMax|$ for all of the studied ARs (computed by replacing $\tauc$ in Equation~(\ref{eq:Nt}) by $|\taucMean|$ and $|\taucMax|$, respectively).
}
\label{fig:histo-Nt}
\end{center}
\end{figure}

Next, using the simple flux-rope model, we convert the measured $\tauc$-values to the number of turns present in half of the emerging torus using the relation
  \BE \label{eq:Nt}
  \Ntc = \frac{\tan \tauc}{2},
  \EE
derived in Appendix B of \inlinecite{Luoni11}. This model has $\tauc$ almost constant during most of the emergence phase (\fig{modN1}) because the flux rope is uniformly twisted.  This is rarely found in observed ARs, as shown in the examples of \figss{10664}{10569}. Then, we compute for the 41 ARs in our set two extreme values, $\NtcMean$ and $\NtcMax$, by replacing $\tauc$ by $\taucMean$ and $\taucMax$, respectively, in  
Equation~(\ref{eq:Nt}). Because of the dependence between $N_{t,c}$, which we use as a measure of the AR twist,  and $\tau_c$ (shown by Equation~(\ref{eq:Nt})) the twist derived in the observed ARs is a monotonically increasing function of $\tau_c$. This relationship is model dependent and can be tested by already developed numerical simulations of flux-rope emergence (see references in \sect{introduction}), 
as both $\tau_c$ and the number of turns can be estimated independently. 

All of the values of $\NtcMean$, except one, are below unity and the median is only $0.2$ (\fig{histo-Nt}). Most ARs have also $\NtcMax$ below unity and the median is only $0.4$.   This shows that the emerging part of the sub-photospheric flux rope typically has a low amount of twist and that highly twisted flux ropes are rare. This is a constraint for models simulating the emergence of flux ropes and, even more, for those which place the flux rope formation at the bottom of the convective zone. Finally, the difference between the histograms of  
$\NtcMean$ and $\NtcMax$ is important as $\tauc(t)$ is typically evolving during the emergence.   This is likely due to a variable twist across the flux-tube (in contrast to our simple model).


\section{Conclusion} 
      \label{sec:conclusions} 

Magnetic tongues, which are attributed to the presence of twist in the flux-tubes that form active regions (ARs), are generally observed during their emergence phase. In this article we analyze how 
frequently tongues are seen in emerging ARs and how their presence affects the evolution of the photospheric-flux distribution and, therefore, the parameters that characterize simple bipolar ARs. With this objective, we select all of the bipolar ARs that were observed emerging on the Sun's surface over eight years, from January 2003 to January 2011. After applying several selection criteria (isolated ARs, low background flux, and emergence close to disk center), required so that magnetic tongues could be observed without distortion, we are left with a set of 41 ARs. From this set, 39 cases were observed by MDI and two by HMI. Then, our set contains examples with high spatial
resolution for which we obtain very similar results as with MDI spatial resolution (as expected, since tongues have length scales comparable to ARs so that they do not require high spatial resolution to be observed).       

Our new method to characterize the presence of magnetic tongues consists in measuring the acute angle, which we call $\tau$, between the photospheric inversion line (PIL) and the direction orthogonal to the main AR bipolar axis (\fig{schema}, \sect{aim}). 
This angle [$\tau$], for our simple emergence model, is a monotonically increasing function of the twist of the flux rope forming the active region and always has the same sign as its magnetic helicity. 
We design an automatic procedure to determine the PIL location, which is based on minimizing the contribution of the magnetic flux that is located on the wrong-sign side of the PIL after automatically subtracting the contribution of the strong-field regions since they are on the right-sign side of the PIL (\sect{pil}). In this sense our numerical procedure can be applied systematically to the full extent of a bipolar AR.  Our method is robust with respect to noise and the presence, or not, of parasitic polarities since it is based on the minimization of an integral function computed over the AR magnetogram. 
Our algorithm is more versatile than the method used by \inlinecite{Luoni11} since in that article the determination of the PIL direction had to be done by selecting 
the portion of the AR having magnetic tongues.  Furthermore, we are computing $\tau$ taking into account the main AR bipolar axis at each time step contrary to what was done by \inlinecite{Luoni11}, where the direction of the bipolar axis was computed at the time of maximum flux, a time which is difficult to defined and needs the user's intervention. Hence our algorithm is more robust and less user- dependent than the one used in \inlinecite{Luoni11}.
Another extension of our procedure is that we can automatically compute the confidence interval within which $\tau$ varies.

The main aim of \inlinecite{Luoni11} was to show that the tongue pattern allows us to determine the sign of the magnetic helicity.  For that purpose, most of the selected ARs were already studied in previous articles and the helicity sign was known. In the present study, our aim is rather to find out how frequent and how marked are magnetic tongues in emerging ARs.   For that purpose we systematically selected, in a time period of eight years, all of the bipolar ARs where the emerging phase could be fully observed close to the central meridian passage.   We found that 63\,\% of the ARs have an average absolute value of $\tau$ [$|\tauMean|$] larger than $20\degree$ (\fig{histo-tau}) showing that a majority of the emerging ARs have clear magnetic tongues that indicate the emergence of a twisted flux-tube (see examples in \figss{10664}{11142}).  Tongues are also present in most of the remaining 37\,\% of cases, but not necessarily during all of the emergence period.  Indeed, only 10\,\% of emerging ARs have a maximum $|\tau |$ lower than $20\degree$ (see \fig{10569}).  Such cases have small magnetic tongues and a careful study of the magnetogram time series is needed to identify or eliminate tongues.  We conclude that the large majority of ARs are formed by the emergence of a twisted flux-tube (or several).  Such emergence is fragmented in the form of a small sea-serpent-like flux tube.  Nevertheless the large-scale pattern of the tongues, at the AR scale, indicates the presence of a coherent flux rope below the photosphere. 

   The azimuthal component of the flux-rope magnetic field is the origin of the observed magnetic tongues.
This extra magnetic flux artificially rotates the magnetic bipole axis, defined by the center of gravity of both AR polarities (in contrast with a real rotation of the flux rope).   To quantify this effect, we investigate how the bipole tilt [$\phi$] evolves during the emergence.  For 83\,\% of the ARs,  with respect to its value when the flux is maximum, the mean change of $\phi$ is below $10 \degree$ (\fig{histo-phi}).  We conclude that in most emerging ARs, the presence of magnetic tongues has a small effect on the AR tilt determined from the magnetograms.  However, some ARs do have a large variation of the tilt, up to $50 \degree$.

   We next estimate the orientation of the flux rope from the measured tilt when its magnetic flux is maximum and the tongues have retracted.  We correct all ARs for the effect of the magnetic tongues on the measured tilt, providing a corrected $\tau $, called $\tauc $. Coherently with the results related to the tilt, $\tau \approx \tauc, $ within $10 \degree$, for most ARs.  However, a fraction of ARs have $|\tauc|$ significantly larger than $|\tau|$, in agreement with a simple model for the kinematic emergence of a uniformly twisted flux rope (\fig{correl-tau}).
     
The measurement of $\tauc $ allows us to estimate the twist by comparison with our flux-rope model.
We derive the equivalent number of turns [$\Ntc $] present in half of the emerging-torus model.  We find a low twist, since half of the emerging ARs have a mean $\Ntc $ below $0.2$ and a maximum $\Ntc $ below $0.4$.  Only 12\,\% of ARs have a maximum $\Ntc $ above one turn.  So, while nearly all emerging ARs show the presence of a flux rope, the amount of twist in the emerging part is low. There are other methods to infer the degree of twist in emerging ARs, which can be more complex than our set of simple bipolar ARs, \eg\ the computation of photospheric electric currents or the current helicity which require vector magnetograms (\opencite{Gosain14,Zhang01,Zhang12}). Our method relies on the determination of a single acute angle during the earliest period of flux emergence and, therefore,  has the advantage of giving a first quantitative estimate of the AR twist directly derived from line-of-sight magnetograms.  

Our result concerning the low twist in the sub-photospheric emerging part of a flux rope is coherent with the estimation of the magnetic-helicity injection found during the emergence phase of ARs and the helicity stored in the coronal field. The photospheric helicity flux has been derived from
the temporal evolution of magnetograms. Its sum over the emergence phase, normalized by the maximum magnetic flux to the second power, provides an estimation of the number of turns (of an equivalent uniformly twisted flux-tube). These estimations range from a few times 0.01 to a maximum value just above 0.1 (\opencite{Nindos03,Liu06,Jeong07,Labonte07,Tian08,Yang09}).  The magnetic helicity contained at a given time in the coronal field can also be estimated from magnetic-field extrapolations of the photospheric magnetograms, using the force-free model that best fits the observed coronal loops. This provides an estimation independent of the helicity injection computed above. The equivalent number of turns typically ranges from 0.1 to 0.2 (\opencite{Demoulin02a,Green02,Mandrini05,Lim07}).   

We conclude that our results, in relation to $\Ntc $ estimated from the tongue shapes, are broadly compatible with the independent estimations of magnetic helicity pointing to a low twist, well below one turn, for the full AR configuration.   These observational results set severe constraints on the numerical models of emerging flux ropes and even more on those that study the crossing of the convective zone by flux ropes formed at its base. 
Although we use a simple model to estimate the twist from $\tauc$, the results of numerical simulations of emergence (see references in \sect{introduction}) can be analyzed as observations and their $\tauc$-values can be directly compared. This would provide a direct test of these numerical results.  
Finally, another extension of this investigation is to analyze the relationship between the number of turns derived from our estimated  $\tauc$-values (average and maximum) and those inferred from force-free models of the coronal magnetic field for the same ARs.

%
\begin{acks}
CHM and MLF acknowledge financial support from the Argentinean grants PICT 2012-0973 (ANPCyT), UBACyT 20020100100733 and PIP 2009-100766 (CONICET). 
CHM and MLF are members of the Carrera del Investigador Cient\'ifico (CONICET). 
MP is a fellow of CONICET.
CHM thanks the Paris Observatory for a one-month invitation.    
\end{acks}


\newpage

\section{Appendix} 
	\label{sec:appendix} 

\begin{center}
\begin{longtable}{C{.6in} C{.5in} C{.5in} C{.5in} C{.5in} C{.5in} C{.5in}}
\caption{The analyzed ARs and their computed parameters. Column one shows the NOAA AR number. Columns two to five list 
the values of  $\tauMean$, $\tauMax$, $\taucMean$, and $\taucMax$; all of these values are expressed in degrees. Finally, columns six and seven, 
indicate the number of turns as defined in the text and computed using $\taucMean$ and $\taucMax$, respectively. All of the parameters are computed during the analyzed AR emergence period.} \label{tab:ars} \\
\toprule[1.5pt]
{NOAA}  &  {$\tauMean~[\deg]$}  &  {$\tauMax~[\deg]$}  &  {$\taucMean~[\deg]$}   &  {$\taucMax~[\deg]$}  &  {$\bar{N}_{\rm t,c}$}  &  {$N_{\rm t,c,max}$} \\
\midrule
\endfirsthead
\multicolumn{7}{c}%
{\tablename\ \thetable\ -- \textit{(Continued.)}} \\
\toprule[1.5pt]
{NOAA}  &  {$\tauMean~[\deg]$}  &  {$\tauMax~[\deg]$}  &  {$\taucMean~[\deg]$}   &  {$\taucMax~[\deg]$}  &  {$\bar{N}_{\rm t,c}$}  &  {$N_{\rm t,c,max}$} \\
\midrule
\endhead
\bottomrule[1.25pt] 
\endfoot
\bottomrule[1.25pt]
\endlastfoot

10268  &  -34  &  -52  &  -59  &  -76  &  -0.8  &  -2 \\
10274  &  32  &  41  &  49  &  58  &  0.6  &  0.8\\
10311  &  -21  &  -35  &  -20  &  -34  &  -0.2  &  -0.3\\ 
10319  &  -8.3  &  -21  &  -10  &  -21  &  -0.09  &  -0.2\\ 
10323  &  33  &  47  &  26  &  38  &  0.2  &  0.4\\ 
10344  &  11  &  41  &  12  &  39  &  0.1  &  0.4\\ 
10349  &  -13  &  -41  &  -15  &  -46  &  -0.1  &  -0.5\\
10381  &  -30  &  -39  &  -31  &  -40  &  -0.3  &  -0.4\\ 
10385  &  29  &  39  &  28  &  39  &  0.3  &  0.4\\ 
10391  &  -33  &  -53  &  -31  &  -51  &  -0.3  &  -0.6\\ 
10415  &  -15  &  -34  &  -20  &  -44  &  -0.2  &  -0.5\\
10440  &  -18  &  -39  &  -19  &  -38  &  -0.2  &  -0.4\\ 
10441  &  -22  &  -35  &  -17  &  -31  &  -0.2  &  -0.3\\ 
10445  &  -22  &  -35  &  -27  &  -38  &  -0.3  &  -0.4\\ 
10456  &  26  &  48  &  28  &  49  &  0.3  &  0.6\\ 
10488  &  7.4  &  15  &  -6.6  &  -25  &  -0.06  &  -0.2\\
10495  &  -47  &  -59  &  -51  &  -63  &  -0.6  &  -1\\
10547  &  -22  &  -41  &  -38  &  -54  &  -0.4  &  -0.7\\ 
10565  &  27  &  56  &  42  &  57  &  0.5  &  0.8\\ 
10569  &  -5.8  &  -23  &  -10  &  -33  &  -0.09  &  -0.3\\ 
10617  &  -7.3  &  -26  &  -6.1  &  -26  &  -0.05  &  -0.2\\
10664  &  33  &  45  &  68  &  81  &  1.2  &  3.2\\ 
10692  &  -20  &  -41  &  -39  &  -59  &  -0.4  &  -0.8\\ 
10727  &  -29  &  -52  &  -40  &  -63  &  -0.4  &  -1\\ 
10747  &  -23  &  -39  &  -6.2  &  -21  &  -0.05  &  -0.2\\ 
10828  &  45  &  68  &  57  &  81  &  0.8  &  3.3\\ 
10837  &  -14  &  -27  &  -11  &  -27  &  -0.1  &  -0.3\\ 
10879  &  -16  &  -25  &  -23  &  -33  &  -0.2  &  -0.3\\ 
10900  &  40  &  63  &  49  &  72  &  0.6  &  1.5\\ 
10955  &  27  &  53  &  21  &  45  &  0.2  &  0.5\\ 
10971  &  23  &  30  &  28  &  33  &  0.3  &  0.3\\ 
10987  &  6.6  &  26  &  4.4  &  25  &  0.04  &  0.2\\ 
11005  &  9.4  &  16  &  15  &  23  &  0.1  &  0.2\\ 
11007  &  -40  &  -66  &  -37  &  -65  &  -0.4  &  -1.1\\ 
11010  &  -22  &  -39  &  -18  &  -35  &  -0.2  &  -0.3\\ 
11024  &  -21  &  -45  &  -21  &  -54  &  -0.2  &  -0.7\\ 
11027  &  -9.9  &  -22  &  -15  &  -42  &  -0.1  &  -0.5\\ 
11043  &  45  &  54  &  47  &  59  &  0.5  &  0.8\\ 
11049  &  -15  &  -36  &  -10  &  -30  &  -0.09  &  -0.3\\ 
11141  &  -24  &  -41  &  -26  &  -43  &  -0.2  &  -0.5\\ 
11142  &  34  &  56  &  28  &  53  &  0.3  &  0.7\\ 
\end{longtable}
\end{center}

 
\bibliographystyle{spr-mp-sola} 
\bibliography{paper_tongues} 
\IfFileExists{\jobname.bbl}{}
{\typeout{}
\typeout{****************************************************}
\typeout{****************************************************}
\typeout{** Please run "bibtex \jobname" to obtain}
\typeout{** the bibliography and then re-run LaTeX}
\typeout{** twice to fix the references!}
\typeout{****************************************************}
\typeout{****************************************************}
\typeout{}
}

\end{article} 
\end{document}